# Local electrochemical characterization of active Mg-Fe materials – from pure Mg to Mg50-Fe composites

Noémie Ott[a,ǂ*], Aurélien Tournier Fillon[a], Oliver Renk[b], Thomas Kremmer[c], Stefan Pogatscher[c], Thomas Suter[a,†], Patrik Schmutz[a,*]

[a] Laboratory for Joining Technologies and Corrosion, Empa – Swiss Federal Laboratories for Materials Science and Technology, Überlandstrasse 129, 8600 Dübendorf, Switzerland

[b] Department Materials Science, Montanuniversität Leoben, Roseggerstraße 12, 8700 Leoben, Austria

[c] Chair of Nonferrous Metallurgy, Montanuniversität Leoben, Franz-Josef-Strasse 18, 8700 Leoben, Austria

* Corresponding authors: noemie.ott@ost.ch (N. Ott), patrik.schmutz@empa.ch (P. Schmutz)

† Deceased on April 25th, 2020 (Dr. T. Suter)

ǂ Current address (N. Ott): Institute for Microtechnology and Photonics, OST, Werdenbergstrasse 4, 9471 Buchs, Switzerland

**Highlights**

- SEN allows assessing the reaction kinetics at the sub-micrometer range
- Direct visualization and assessment of intermetallic phases nobility is provided
- Insights about nanoscale galvanic coupling within an intermetallic particle can be retrieved
- Surface reactivity of Mg50-Fe composites is dictated by their Mg phase composition
- Changes in phase spacing and composition are observed depending on HPT processing

**Abstract**

This study demonstrates the applicability of the scanning electrochemical nanocapillary (SEN) technique to characterize the local surface reactivity of active systems, such as Mg-based materials. Owing to its confined electrolyte configuration, one undeniable strength of the method is to provide with unprecedented resolution direct visualization and assessment of the presence, distribution and nobility of different phases. High lateral resolution open-circuit potential (OCP) scans on single Fe-rich particles in Mg confirms that these particles serve as local cathodes while evidencing enhanced surface activation at the interfacial area between the particle and the Mg matrix. Valuable insights about nanoscale galvanic coupling within an intermetallic particle can therefore be retrieved, which are otherwise not accessible. On more complex systems, such as Mg50-Fe composites, the SEN technique allows individual assessment of the reactivity of the different microscale phases. By combining OCP scans and local potentiodynamic polarization measurements, we reveal that changes in surface reactivity and stability of Mg-rich phases in these composites are directly correlated to their different microstructure, i.e. phase spacing and composition, which are intrinsically linked to their processing parameters. The SEN technique is therefore an excellent tool to help us refine our mechanistic understanding of initial stages of corrosion in heterogeneous materials.

**Graphical abstract**

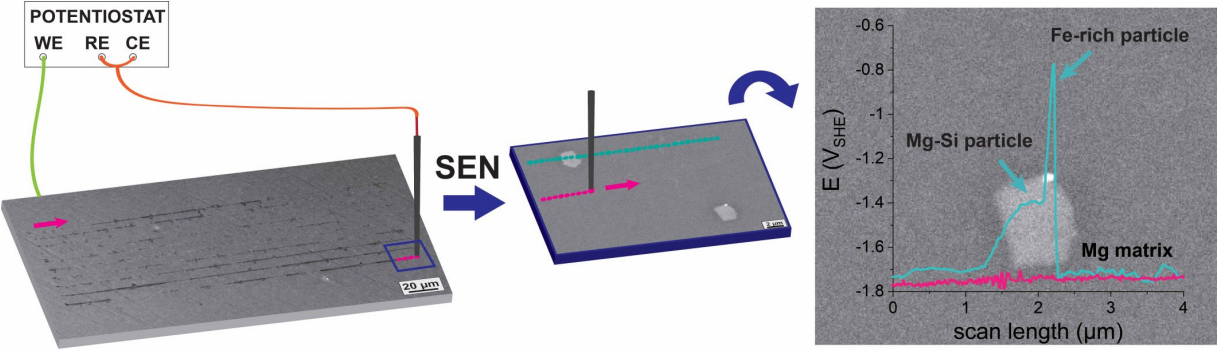

# 1. Introduction

With the current challenges faced by our societies in terms of resources, sustainability and health, magnesium and its alloys have received increasing attention for many technical applications ranging from lightweight transportation to hydrogen-based energy storage and biodegradable implants. Although their high strength-to-weight ratio, recyclability and biocompatibility/adaptability make them particularly attractive, their limited formability at low temperature, low inherent strength and poor resistance to corrosion still limits their popularization.

Since pure Mg does not provide sufficient strength, it requires alloying for most applications, facing the paradox that a large number of elements have no or very limited solubility in Mg [1]. Moreover, Mg is anodic to most elements and therefore, highly susceptible to micro-galvanic coupling, originating from the potential difference between second phase particles formed by alloying elements or impurities and the Mg matrix, causing rapid degradation [2–6]. Impurities, such as Fe, are particularly detrimental and thus the concept of tolerance limits, above which the corrosion rate drastically increases, has been introduced already in the 1940s [2]. Hanawalt *et al* [2] have empirically determined an Fe tolerance limit of 170 ppm (0.017 wt.%), which corresponds to the eutectic composition in the calculated Mg-Fe phase diagram [7]. Above that value, stable Fe-rich phase forms during solidification, leading to high corrosion rates. The tolerance limit is however not immutable. The presence of other impurities as well as the alloy processing can contribute to lower it [4,7–10].

Therefore, in recent years, new strategies for alloy design have emerged with, for instance, the development of ultra-high purity magnesium with trace amount of Fe [11], which exhibits extremely low corrosion rates in common chloride containing solutions [12–14]. The presence of impurities is however hardly unavoidable in an industrial large scale manufacturing process and subsequently, other strategies involve controlling the Mg alloy microstructure through the use of thermo-mechanical processing [7,8,15,15–18]. In particular, non-conventional processing routes involving severe plastic deformation (SPD), such as high pressure torsion (HPT), are attractive for the preparation of Mg alloys for hydrogen storage in bulk form [19–22] or for biomedical applications [23–28].

Undeniably, the microstructure and in particular the presence of intermetallic phases is determining for the corrosion performance of Mg alloys. When exposed to an electrolyte, the differences in nobility between the microstructural features dictate the surface reactivity of the alloys. Beyond simply determining the microscale galvanic coupling between the intermetallic phases and the Mg matrix, assessing the reaction kinetics at the interfacial area is key to an improved understanding of Mg corrosion processes. In particular, defining a "realistic" environment-specific corrosion rate is critical for biomedical applications, and also in the evaluation of the efficiency of corrosion protection strategies [29,30].

Local electrochemical characterization can provide insightful mechanistic information by individually assessing the behavior of the different microstructural features in electrolytes. One particularity of Mg corrosion is that macroscale electrochemical measurements do not necessarily reflect the real corrosion behavior of Mg alloys and can lead to an underestimation of the corrosion rates compared to direct measurements, such as hydrogen collection. This is due to the occurrence of anomalous hydrogen evolution (HE), referring to the large increase in HE rate with increasing applied anodic current or potential [4]. In the last decade, the use of scanning probe techniques significantly contributed to new insights into Mg corrosion mechanisms [14,31–33,33,33–36,36–42]. For instance, scanning vibrating electrode technique (SVET) has been employed to track changes in localized current densities on corroding Mg surfaces, allowing to identify and differentiate sites with cathodic and anodic activity. Notably, on unpolarized Mg surfaces, Williams *et al.* [32,33] evidenced a cathodic activation of the previously anodically attacked surface. SVET, as well as scanning electrochemical microscopy (SECM), have moreover provided valuable information in the study of anomalous HE on anodically polarized Mg surfaces [37,42–44]. More recently, they have further evidenced that in addition to HE, significant amount of oxygen reduction reaction occurs during Mg corrosion as theoretically expected, in particular for high purity Mg alloys [14,36]. These techniques have therefore shown great potential to investigate Mg corrosion mechanisms. However, one of their main limitations is that their resolution is strongly dependent on electrolyte nature and conductivity as well as probe-sample distance. These factors can be drastically affected by the significant surface modification occurring during Mg corrosion as well as possible interference resulting from hydrogen formation [45,46]. The surface reactivity at the interfacial area between intermetallic phases and Mg matrix can be obtained with high lateral resolution information by scanning Kelvin probe force microscopy (SKFPM) [8,38,47,48], but the method does not provide insights on reaction kinetics in bulk electrolyte.

Therefore, the present study explores the applicability of the scanning electrochemical nanocapillary (SEN) method to investigate local corrosion behavior of Mg-based materials. The SEN technique is based on the electrochemical scanning capillary microscope developed by Böhni H. *et al* [49–51] for nanolithography. A borosilicate glass nanocapillary filled with electrolyte acts as a probe and is scanned, or rather hopped, over the sample surface, following an intermittent contact principle. After the approach, the probe remains on a point for some time (from a few milliseconds to seconds) to allow meniscus-confined electrochemical measurements. The probe is then retracted and moved to the next point. Compared to the other scanning probe techniques, only the area underneath the probe is exposed to the electrolyte, which prevents the droplet to be dragged from one measurement point to the next one. SEN measurements consequently allow a direct correlation of the measured electrochemical response to specific micro- or nanostructural features.

The purpose of this study is to assess at the sub-micrometer range, the influence of anodic and cathodic areas on the initial stages of corrosion processes of pure Mg as a function of impurity content and processing. The observations are validated on model Mg50-Fe composites produced by high pressure torsion (HPT). In this case, the processing conditions proved to be decisive for the surface reactivity and phase stability of the Mg-rich phases of the studied composites.

## 2. Experimental

### 2.1. Materials

A commercially pure (CP) magnesium grade (99.9%, as rolled, Goodfellow) was compared to a high purity (HP) magnesium grade for investigating the effect of impurity content on the corrosion behavior of pure Mg. High purity Mg substrates were produced by permanent mould direct chill casting (Helmholtz Zentrum Geesthacht) with their impurity content remaining lower than their tolerance limit – typically Fe <170 ppm. High purity Mg was studied in the as-produced state and after heat-treatment (HT) at a temperature of 525 °C (798 K), held for 24 h under constant Ar flow (Ar 6.0, PanGas, ~200 ml/min). The chemical compositions of the two pure Mg grades, as given by the manufacturer, are listed in Table 1.

Table 1: chemical composition of the two studied pure Mg grades (wt. %)

| wt. % | Mg | Al | Mn | Fe | Zn | Cu | Ni |
|---|---|---|---|---|---|---|---|
| **pure Mg – high purity (HP)** | bal. | 0.0070 | 0.0080 | 0.0045 | 0.0100 | | <0.0010 |
| **pure Mg – commercial purity (CP)** | bal. | 0.0070 | 0.0170 | 0.0280 | 0.0050 | 0.020 | <0.0010 |

Model Mg50-Fe composites (vol. %) were also investigated as an example of high-Fe containing materials. Coarse Mg50-Fe composites were produced by cyclic high-pressure torsion (HPT) at room temperature (10 degrees twist angle for 20 cycles) [52], while nanostructured Mg50-Fe composites were produced by monotonic HPT at 300 °C (573 K, 15 rotations). The applied nominal pressure was in both cases 7.8 GPa. Disks of 8.0 mm dimeter and about 1 mm height were obtained. Note that due to the radial strain dependence in torsion, HPT processing leads in case of composites and cyclic processing to some microstructural heterogeneities across the disk diameter. In this study, only the intermediate region of the sample is considered.

All the samples were ground to a 4000 silicon carbide paper finish and subsequently polished down to 1 μm in ethanol. For microscopic observations, final polishing was performed with diluted sol-gel alumina suspension solution (MasterPrep, 0.05 μm, Buehler). The samples were then ultrasonically cleaned in ethanol and dried with Ar.

## 2.2. Scanning electrochemical nanocapillary (SEN) characterization

Local surface reactivity, i.e. thermodynamic and kinetic information, of the pure Mg grades and the Mg50-Fe composites, was characterized using the scanning electrochemical nanocapillary (SEN) technique presented in Figure 1a. Based on meniscus-confined electrochemical measurements, a borosilicate glass nanocapillary filled with electrolyte acts as probe and is scanned, or rather "hopped" over the sample surface. In the paper, the term "scanned" is used, but for each point, a full sequence of controlled approach, measurement and retractation is performed before moving to the next point. Figure 1b illustrates the operational principle.

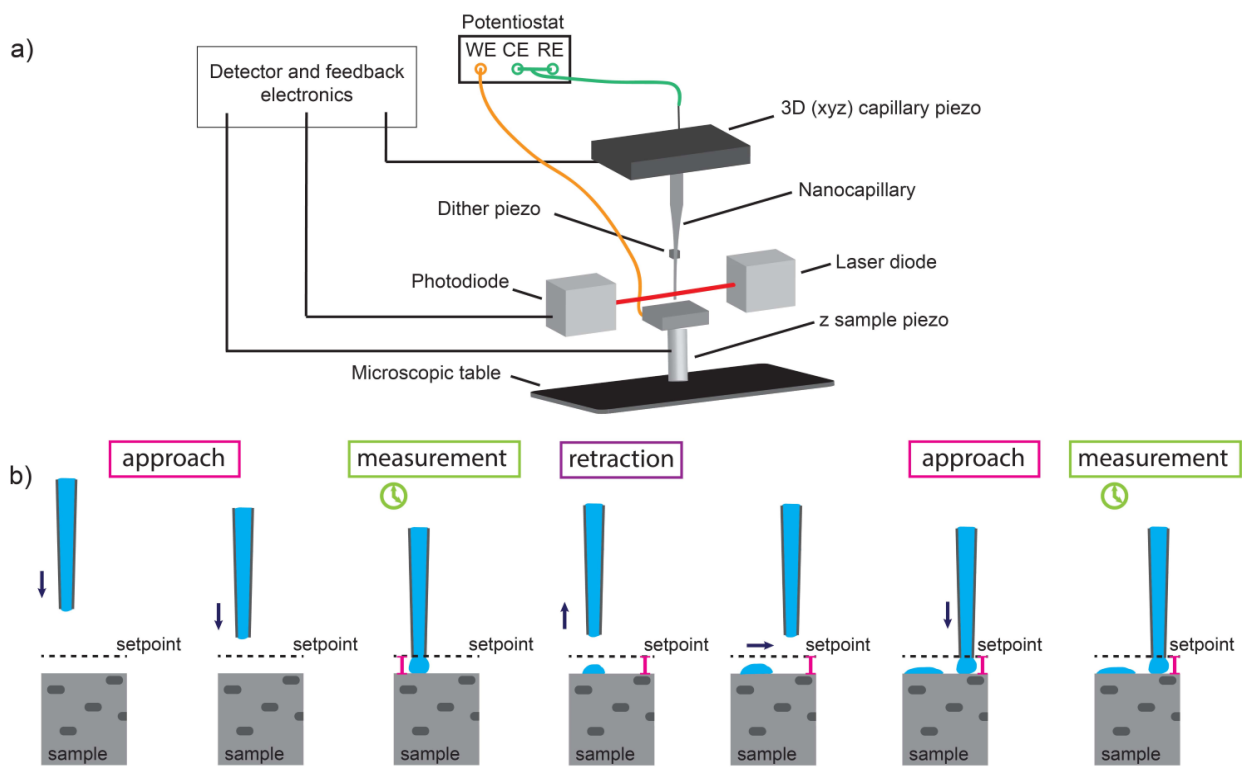

Figure 1: a) Schematic description of the scanning electrochemical nanocapillary (SEN) setup. b) details of the SEN intermittent contact principle on flat samples based on meniscus-confined electrochemical measurements. After the approach, the probe remains on a point for some time (from a few milliseconds to seconds) to allow electrochemical measurements. The probe is then retracted and moved to the next point.

In this study, probes with an opening diameter of 40 nm filled with 0.1 M HCl were used. The probes are obtained by pulling borosilicate glass capillaries (OD: 1.0 mm, ID: 0.58 mm, with filament, Science Products) in a one-stage process with a laser-based micropipette puller (P-2000, Sutter Instruments). The capillary tip dimension was checked with STEM-in-SEM imaging, using a FEI Nova NanoSEM 230 field emission gun scanning electron microscope (FEG-SEM) equipped with a STEM II – FP 6903/06 annular scanning transmission electron microscopy (STEM) detector (Figure S1). Before each scan, the resonance frequency of the probe is determined but typically lies in the 40 – 80 kHz range and only probes with a Q-factor above 100 are considered (Figure S2). The distance between the capillary tip and the sample surface is controlled by a proportional integral derivative (PID) feedback loop implemented on the sample piezo which ensures that the distance remains constant during the measurement duration on a point, i.e. contact with the electrolyte droplet, and so during the whole duration of the scan. The height variation of the z sample piezo (Figure 1) is therefore directly correlated to the sample topography.

As the tip of the nanopipette is brought close to the substrate (approach), a liquid meniscus is formed. Consequently, simultaneously to tracking the topography (Figure S3), local electrochemical measurements can be performed, using a high-resolution Jaissle IMP 83 PCT-BC potentiostat, with the nanocapillary serving as electrochemical cell. In this study, a two-electrode setup was used with the sample acting as working electrode and a Pt wire being the quasi-reference electrode/counter electrode (RE/CE). The potentials are expressed versus the standard hydrogen electrode ($E_{Pt}$ = 605 mV$_{SHE}$). The SEN ability to measure microscale potential changes was demonstrates on a PFKPFM-SMPL reference sample (Bruker) and is reported in Figure S4.

Potentiodynamic polarization curves were performed on a selected area with a scan rate of 10 mV s$^{-1}$. The current measured was not corrected for the area. The exposed area is defined by the meniscus/droplet formed on the sample surface and highly depends on the sample surface wettability and reactivity. Considering a passive surface, the exposed area is not expected to change during electrochemical measurements and can be estimated by post-mortem scanning electron microscopic imaging. On the contrary, in the case of Mg materials exposed to chloride-containing electrolytes, the effective surface is expected to drastically increase as soon as active dissolution starts, which can lead to an estimation mismatch (Figure S5).

Complementary "fast" line scans on Mg alloy surfaces were conducted at open-circuit potential (OCP). The local potential changes displayed are obtained by averaging the potential values over the last 25 ms.

### 2.3. Further materials characterization

Scanning electron microscopic (SEM) analysis of the Mg sample surfaces before and after SEN measurements was performed using a Hitachi S3700N instrument equipped with an EDAX/Ametek Octane Pro energy dispersive X-ray (EDS) detector.

Microstructural analysis of Mg50-Fe composites was performed using scanning transmission electron microscopy (S/TEM) combined with energy-dispersive X-ray spectroscopy (EDS). Imaging was performed using a Thermo Fisher Scientific Talos F200X G2, operating at 200 kV, fitted with an on-axis bright-field (BF) detector and a high-angle annular dark-field (HAADF) detector. TEM samples were prepared by focused ion beam (FIB) using a Zeiss 1540XB CrossBeam. Environmental scanning Kelvin probe force microscopy (AFM-SKPFM) was performed using a Bruker ICON 3 atomic force microscope at a scan rate of 0.120 Hz. A PFQNE-AL probe was employed for Volta potential measurements. The reliability of the probe used was assessed by measuring the Volta potential of pure Ni (99.0 %), pure Fe (99.5 %) and pure HP Mg. The potentials are expressed vs the Volta potential of pure Ni reference as introduced in [53].

Structural characterization of Mg50-Fe composites was performed via X-ray diffraction (XRD) in a Bruker D8 diffractometer in Bragg Brentano geometry using Cu Kα radiation and a Ni filter.

The macroscale corrosion behavior of pure Mg materials was estimated by volumetric hydrogen collection measurements. Pure Mg samples were placed in a beaker containing 1.0 M NaCl at an initial pH value of 6.0. A measuring cylinder also filled with this unbuffered NaCl solution was placed directly on top of the sample in order to collect all the hydrogen gas produced as a result of the dissolution process. The volume of $H_2$ produced by the cathodic reduction is proportional to Mg dissolution and thus, can be directly used to derive the Mg corrosion rate. Complementary immersion tests followed by element-specific wet chemical analysis were performed in unbuffered 0.15 M NaCl at an initial pH value of 5.2 (Suprapur grade). Aliquots were sampled every 10 min during 180 min and subsequently diluted five or ten times with a 1% nitric acid electrolyte (Suprapur grade). The analysis of the solutions was conducted using an Agilent 7500ce inductively coupled plasma mass spectrometer (ICPMS). The calibration was performed using matrix-matched blank solution and multi-element standard solutions in a concentration range from 0.05 µg L$^{-1}$ to 5000 µg L$^{-1}$. An internal standard solution was added online to correct for non-spectral interferences. Only concentrations above the background equivalent concentration (BEC) were taken into account. The dissolved amount of each element was corrected for the initial weight of the sample and is reported as µg g$^{-1}$. The immersion tests were repeated three times for both purity grades.

Electrochemical assessment was performed by means of potentiodynamic polarization curves in 0.1 M NaCl (initial pH value of 6.0) at a scan rate of 0.5 mV s$^{-1}$, using the microcapillary technique [54–57]. A microcapillary with a tip diameter of 100 µm (exposed area: 0.8 mm$^2$) coated with silicon – to ensure a tight sealing – was chosen as micro-electrochemical cell, allowing several measurements on a given sample for measurement consistency and repeatability. Pure Mg samples acted as working electrode, a saturated calomel electrode (SCE) was used as reference electrode and a Pt wire as counter electrode. The polarization measurements were performed in a single upward scan and the potential scan was started in the cathodic region to minimize surface dissolution. The measurements were repeated three times and showed very good reproducibility.

## 3. Results and discussion

## 3.1. Microstructure and macroscale corrosion behavior of pure Mg materials

In all three investigated pure Mg materials (CP, HP and HP-HT), second phase particles were observed (Figure S6). Their size, composition and distribution however, depend on the processing parameters and impurity content. In commercially pure Mg (CP), the impurity content is higher than the tolerance limit and therefore, according to thermodynamic calculations, Fe becomes immiscible in Mg and precipitates as second phase in the Mg matrix. As shown in Figure 2, two types of precipitates were identified: $Mg_2Si$ particles and Fe- and Mn-rich particles. The corresponding EDX analysis is reported in Figure S7. The particle size ranges from 200 nm to a few µm. They are randomly distributed within the sample, with a higher concentration of particles in some area compared to others. On the other hand, high purity Mg (HP Mg) has an impurity content lower than the tolerance limit. HP Mg in the as-processed condition exhibits large grains, with a few µm-sized, randomly distributed $Mg_2Si$ particles. Some of them are decorated with tiny Fe particles, as shown in Figure 2. Heat-treatment of HP Mg at 525 °C for 24 h results in smaller grains but also in segregation of impurity elements at the grain boundaries (Figure S6). In addition, a larger number of $Mg_2Si$ particles decorated with Fe and Mn are found within the grains.

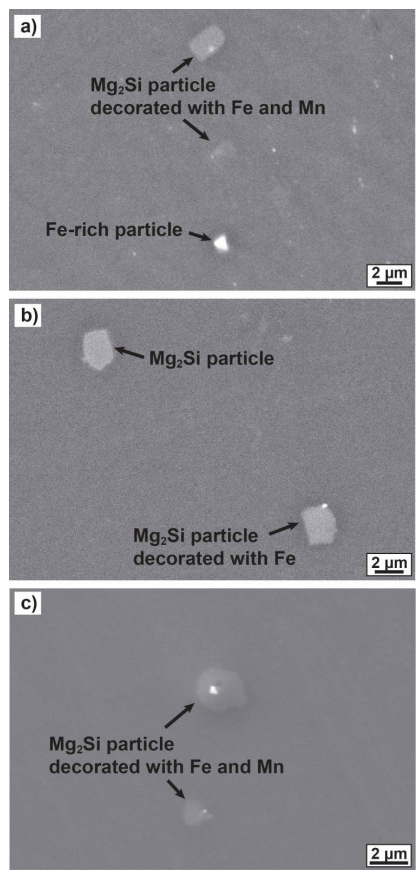

Figure 2: SEM images of particles found in a) commercially pure (CP) Mg specimens and high purity (HP) Mg specimens b) as-produced and c) after heat-treatment for 24 h at 525 °C (HP-HT).

The macroscale corrosion rates were first estimated by volumetric hydrogen gas collection measurements and immersion tests. As expected, CP Mg exhibits much higher dissolution rates than HP Mg. This is clearly evidenced by its higher hydrogen evolution and Mg dissolution rates, as shown in Figure S8. It is also worth noting that element impurities were only detected for CP Mg (Figure S9). Zn follows the dissolution behavior of Mg at a similar ratio to the one of the bulk, which confirms the presence of this element in solid solution. On the other hand, notable amount of Mn related to its initial bulk content is quickly detected in solution, reflecting a preferential Mn dissolution while the dissolution of Fe is delayed and remains relatively low compared to the other alloying elements and impurities (Figure S9). This tends to indicate an additional enrichment of Fe on the corroded CP Mg surface as Fe-remnants and/or due to Fe-redeposition. This accumulation of Fe under or in the corrosion product film can contribute to the observed enhanced cathodic activity under open-circuit conditions [58–61]. Interestingly, the hydrogen collection measurements reported in Figure S8 show that heat-treated HP Mg (HP-HT) has a slightly more active surface than HP Mg in the as-produced condition. Microstructural changes, in particular related to the presence and distribution of intermetallic particles, prove to affect the corrosion performance of Mg alloys, although Figure S8 undeniably shows that chemistry – two orders of magnitude difference between CP and HP materials – dominate over processing effects.

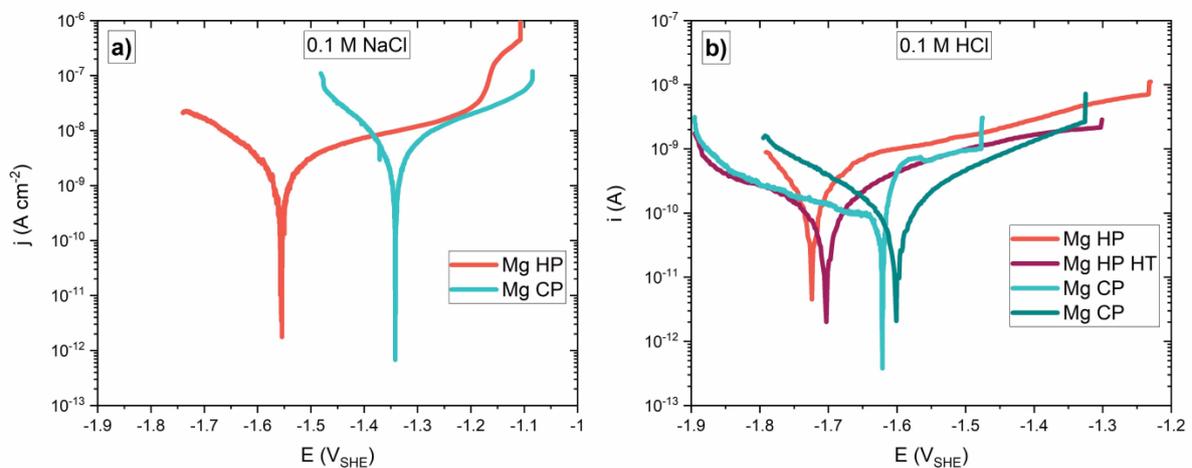

Figure 3: Potentiodynamic polarization curves performed on different pure Mg materials a) at a scan rate of 0.5 mV s$^{-1}$, using the microcapillary technique with a 100 µm diameter microcapillary filled with 0.1 M NaCl and b) at a scan rate of 10 mV s$^{-1}$, using the SEN technique with a 40 nm probe filled with 0.1 M HCl. Note that the current is corrected for the area in the case of the microcapillary technique (area: 0.8 mm$^2$) but not for the SEN measurements.

Potentiodynamic polarization curves have then been performed, using the microcapillary and SEN techniques. Although a shift towards more anodic open-circuit potential (OCP) values is observed for CP Mg compared to HP Mg, both OCP values correspond to an active Mg surface covered with a non-protective oxyhydroxide layer. Figure 3 moreover evidences that even at the micro- and sub-microscale, the catalytic activity for hydrogen evolution reaction (HER) biased the measured anodic branch.

Therefore, no difference in the anodic current response can be detected between both Mg purities even if the corrosion rate determined by other methods and the final level of degradation are completely different. Note that in Figure 3b, the current measured was not corrected by the area. Since active dissolution occurs (Figure S8), the effective surface area increases during the potentiodynamic polarization, even if they only last one minute.

### 3.2. Local surface reactivity of pure Mg alloys - influence of Fe as impurity

A different approach was therefore implemented, aiming at a direct visualization and assessment of local nobilities on pure Mg alloy surfaces. The presence of intermetallic phases in pure Mg is expected to induce changes in surface reactivity in terms of oxide stability and electrochemical response. As previously discussed, Fe plays a determining role in the corrosion of Mg, creating nano- and microscale galvanic coupling, which can result in the initiation of localized corrosion. It is therefore expected that the surface reactivity is significantly enhanced in the vicinity of the Fe-rich particles, in particular at the interfacial area between the intermetallic phase and the Mg matrix. SEN line scans were thus performed under open-circuit conditions to investigate more systematically these changes in local surface reactivity. Figure 4 reports typical OCP scans for CP Mg, HP Mg in the as-produced condition and after heat-treatment for 24 h at 525 °C (HP-HT), using a 40 nm probe filled with 0.1 M HCl. The OCP baseline for Fe is derived from SEN scans performed on pure Fe (Figure S10).

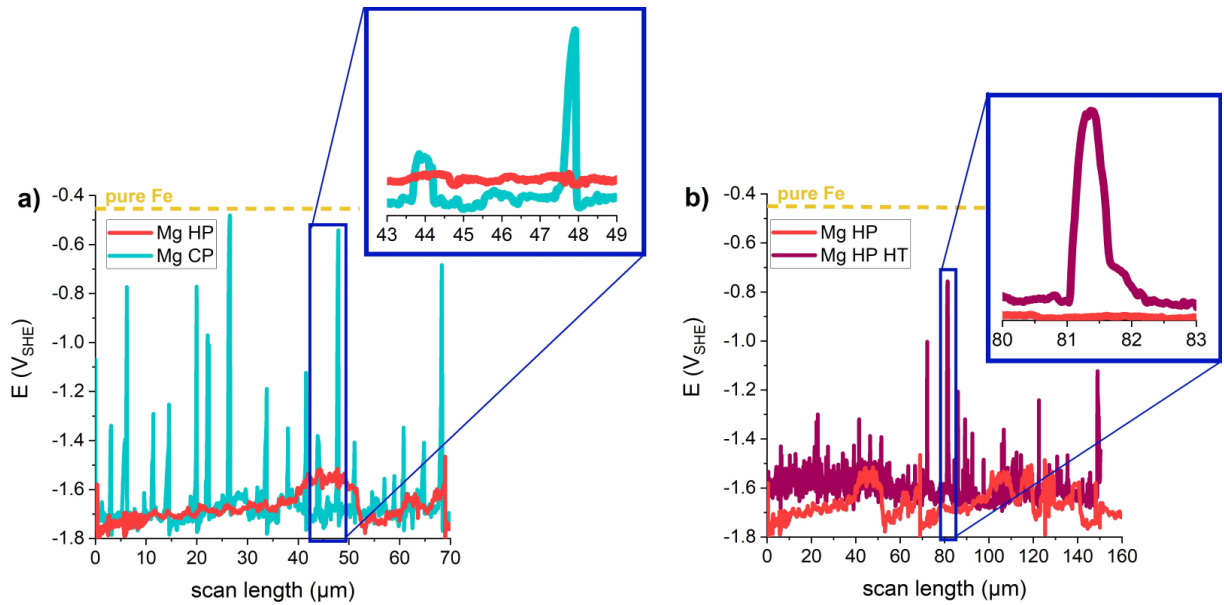

Figure 4: Typical SEN OCP scans for a) high purity Mg (HP) and commercially pure Mg (CP) and b) HP Mg in the as-produced condition and after heat-treatment (HT), using a 40 nm probe filled with 0.1 M HCl. The SEN line scans were performed with a step size of 400 nm, a scanning rate of 500 nm s$^{-1}$ and a measurement duration of 100 ms on each point. The local potential changes displayed are obtained by averaging the potential values over the last 25 ms. Note that the scan length is larger for HP Mg-HT.

The OCP of HP Mg surfaces remains rather constant at a value of -1.7 $V_{SHE}$. This value corresponds to an active Mg surface covered with an oxyhydroxide layer, indicating that despite the use of 0.1 M HCl as electrolyte, staying 100 ms on a point is not enough to fully remove the air-formed oxides and activate the surface. Small scattering is observed and can be attributed to microstructural changes, e.g. grain boundaries and the presence of $Mg_2Si$ intermetallic phases. In comparison, while the potential measured on the Mg matrix for CP Mg and HP Mg HT is similar to the values measured on HP Mg, potential peaks reaching up to -0.45 $V_{SHE}$ are additionally detected. Caption enlargement in Figure 4 indicates that the peak lateral width lies between 500 nm and 2 µm for CP Mg and a couple of micrometers for HP Mg-HT, respectively, which corresponds to the size of the different intermetallic phases identified in these alloys (Figure 2 and Figure S7). As would be expected, these intermetallic phases possess a potential more noble than pure Mg. The potential difference reflects the particle composition, which in SEN OCP scans nicely translates in changes in peak shape and height. When scanned over a particle containing mainly Fe, the potential value increases towards more noble values, close to the one of pure Fe while in comparison, the potential measured for $Mg_2Si$ particles remains close to -1.4 $V_{SHE}$. Smaller potential difference will be obtained when the Mn content of the particles increases. Therefore, less noble spots, in number as well as in potential difference, are measured on HP Mg-HT than on CP Mg. This is also reflected by a clear difference in their corrosion rates (Figure S8).

The SEN technique seems therefore very useful to assess local nobility on heterogeneous material surfaces. This was further demonstrated by SEN OCP scans on single intermetallic phases, namely on an Fe-rich particle (CP Mg surface) and an Fe-decorated $Mg_2Si$ particles (HP Mg HT surface). Note that for these scans, a smaller probe with a nominal diameter of 15 nm was used to achieve better lateral resolution.

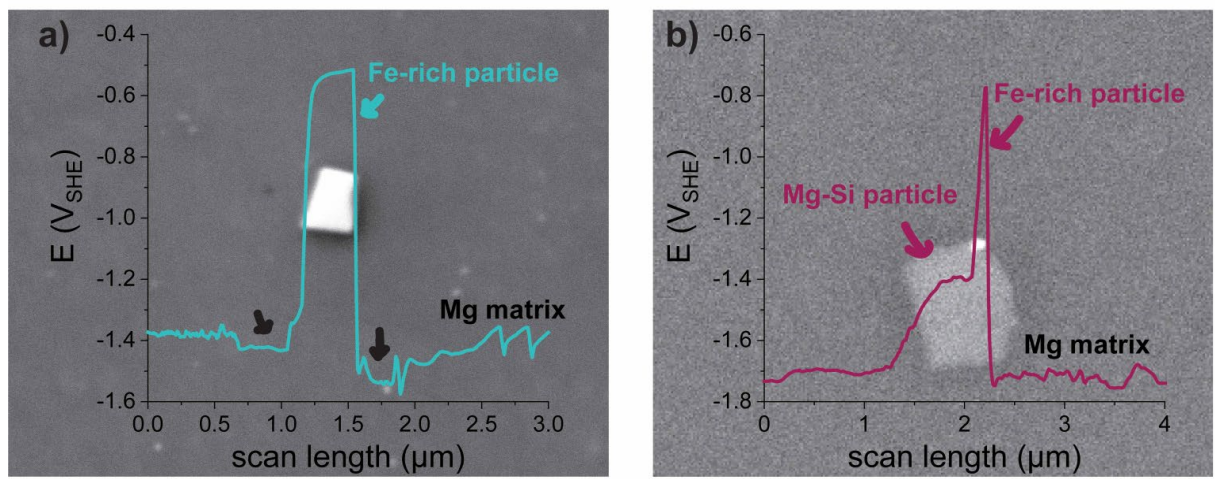

Figure 5: SEN OCP scans on single intermetallic phases: a) Fe-rich particles on CP Mg surface and b) Fe decorated $Mg_2Si$ particles on HP Mg-HT surface, using a 15 nm probe filled with 0.1 M HCl. The black arrows indicate the small potential drop at the particle/Mg matrix interface. The SEN line scans were performed with a step size of 60 nm, a scanning rate of 500 nm s$^{-1}$ and a measurement time of 100 ms on each point. The local potential changes displayed are obtained by averaging the potential values over the last 25 ms.

Figure 5 shows with unprecedented resolution the surface reactivity of single intermetallic phases when locally exposed to 0.1 M HCl. On Fe-rich particle, the OCP value swiftly increases to -0.5 $V_{SHE}$ and remains at this value across the length of the particle. Once the particle is passed, the OCP value drops back to -1.4 $V_{SHE}$, as expected. It is however worth noting the small potential drop at both particle/Mg matrix interfaces (black arrows in Figure 5a), which indicates enhanced Mg surface activation in the vicinity of the Fe-rich intermetallic phase. On the other hand, Figure 5b clearly evidences another nanoscale galvanic coupling within the Fe-decorated $Mg_2Si$ particles with the potential spiking when scanning over the Fe-rich part of the particle. Nevertheless, due to the lower reactivity of the $Mg_2Si$ particles and possibly the lower Fe content in solid solution in the Mg matrix of HP Mg-HT, the OCP of the HP Mg-HT matrix is lower than the one of the CP Mg. No potential decrease at the interfacial area is observed, a different situation compared to Fe-rich particles (Figure 5a) where local Fe depletion around Fe-rich particles is to be expected.

These findings demonstrate that one undeniable strength of the method is the ability to visualize and assess the presence, distribution, and nobility of the different intermetallic phases. Moreover, not only microscale galvanic coupling between noble particles and Mg matrix can be identified but nanoscale galvanic coupling within particles can also be revealed as well as element depletion effects around intermetallic particles. This information is particularly important for Mg or Al alloys, highly sensitive to localized corrosion for whose initiation, the reactivity of the interfacial area between secondary phases and matrix plays a determining role.

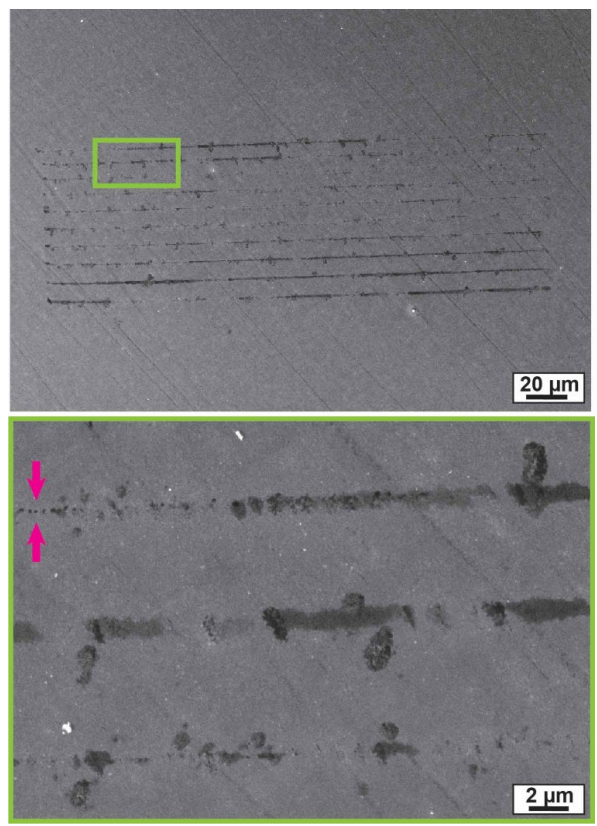

Figure 6: Secondary electron (SE) SEM image after SEN line scans on CP Mg surfaces, using a 40 nm probe filled with 0.1 M HCl. The SEN line scans were performed with a step size of 400 nm, a scanning rate of 500 nm s$^{-1}$ and a measurement time of 100 ms on each point. The distance between each line scan is 5 µm.

In addition to the analysis of specific microstructure/phase reactivity, information about corrosion propagation can be retrieved from SEM observations of SEN line scans, since such observations are typically performed a few hours after the measurements, i.e. after each spot has been exposed to the aggressive HCl environment. Figure 6 shows a secondary electron (SE) image of a scanned area, consisting of ten 250 µm long scan lines (i.e. 625 exposed spots), separated by a distance of 5 µm. Despite the aggressive electrolyte used, only localized dissolution of the oxyhydroxide layer – and no surface activation – was assumed based on the OCP values measured during SEN scan. Disparity in local surface damage (e.g. Figure 6) then directly relates to different surface reactivity of the oxyhydroxide layer. If the surface is not "sufficiently" activated and no lateral corrosion propagation occurs, local oxide destabilizing only leads to dot traces with a diameter of about 200 nm, separated by a distance of about 400 nm (pink pointers in Figure 6). Note that this undeniably proves that the droplet is not dragged from one point to the next one. It further provides direct evidence of the effective exposed droplet area when no active dissolution occurs. Nevertheless, on corrosion susceptible areas, corrosion initiates and propagates, resulting in large trace – up to a few µm wide – and spread corrosion products by the time of the SEM analysis. By comparing SE and backscattered electron (BSE) images, additional insights about surface and in-depth corrosion propagation can be obtained, as shown in Figure S11. These findings demonstrate the applicability of the SEN technique to investigate local corrosion behavior – from initiation on air formed oxy-hydroxide to propagation when stable activation state can be reached – of heterogeneous materials and in particular Mg. SEN OCP line scans allow tracking local surface reactivity while subsequent SEM analysis provides valuable information about oxide destabilization and corrosion initiation/propagation, resulting in the identification of the most corrosion susceptible areas of a material. The SEN technique consequently, is a promising tool to deepen our mechanistic understanding of localized corrosion initiation processes in high-performance alloys.

### 3.3. Influence of Mg50-Fe composite microstructures on their local surface reactivity

The Mg reactivity and influence of Fe distribution has further been investigated using the SEN technique on a Mg-Fe composite containing a very large amount of Fe. The Mg50-Fe composites considered have been developed for biomedical applications. The use of high pressure torsion (HPT) processing to produce these composites allows combining the properties of the individual metals, i.e. high strength and higher corrosion resistance of Fe while keeping the degradation process of Mg, and overcoming some of the technical limitations of conventional processing routes [28]. Processing parameters were selected to obtain two very different composite structures. As shown in Figure S12, cyclic HPT at room temperature leads to a coarse globular microstructure, with micrometer-sized pure Mg and Fe phases while monotonic HPT at 300 °C results in a significantly refined, nanostructured lamellar microstructure. The two composites are therefore referred to as "coarse" and "nanostructured" Mg50-Fe composites hereafter, respectively.

Detailed scanning transmission electron microscopic (STEM) analysis of the two composites evidences that their different processing conditions induce not only changes in the composite structure, i.e. phase spacing, but also in the phase composition. In the coarse Mg50-Fe composite, as would be expected from the processing conditions, the two elements remain in separate phases. A transition region containing both Fe and Mg can nevertheless be found at the interface between the two phases, as shown in Figure 7a. This can be related to the extension of the phases beneath the observed surface of the TEM sample.

On the other hand, high angle annular dark-field (HAADF) images of the nanostructured Mg50-Fe, reported in Figure 7b, reveal the presence of Fe nanoparticles in the Mg phase, resulting most probably from incorporation of lamella fragments, which can occur during monotonic HPT processing. Both composites present oxygen segregation at the grain boundaries that can also affect the composite properties. Notably, regardless of the processing conditions, the composite X-Ray diffraction (XRD) fingerprint only shows intensity reflections corresponding to pure Mg phase in a hexagonal close-packed (hcp) structure and pure Fe phase in a body-centered cubic (bcc) structure (Figure S13). This points out some of the XRD limitations to characterize such complex systems with regards to the distribution of the various elements in the composite.

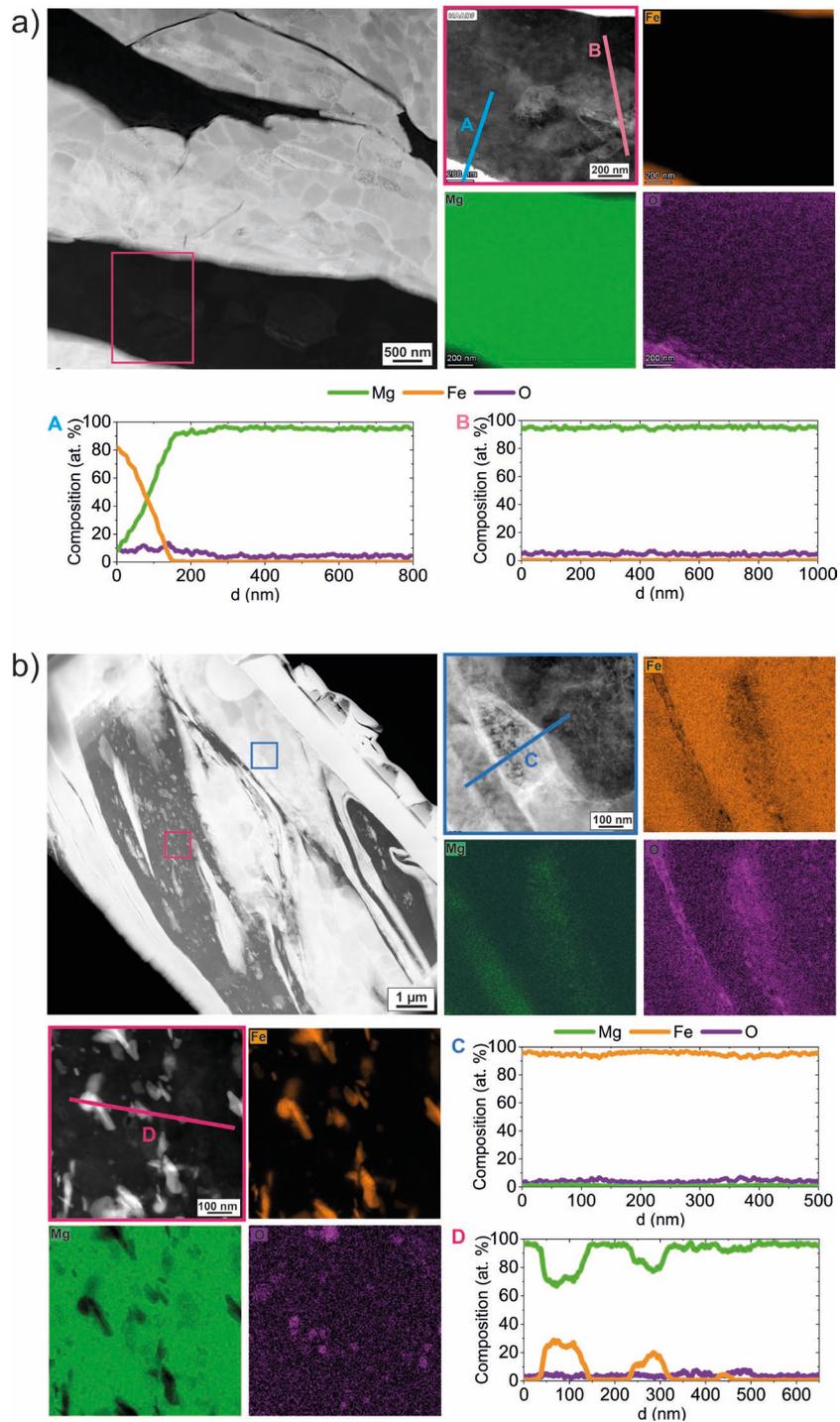

Figure 7: High-angle annular dark-field (HAADF)-STEM images and energy-dispersive X-ray spectroscopy (EDX) maps of a) coarse and b) nanostructured Mg50-Fe composites.

As anticipated, the different microstructures, i.e. phase spacing and composition, of the two Mg50-Fe composites lead to very different macroscale corrosion behaviors [28]. As expected for an Fe phase fraction of 50 vol.%, the Mg phase in the coarse Mg50-Fe composites actively dissolves at high rate upon immersion. Interestingly, nanostructured Mg50-Fe composites present very low dissolution rates as assessed by hydrogen evolution measurements [28] and show little degradation, indicating that, for similar nominal Fe content, the composite microstructure plays a decisive role in their corrosion resistance. To refine the mechanistic understanding of the composite corrosion processes, SEN OCP scans were performed, since they allow to individually assess the reactivity of the two phases. Note that HPT processing leads to some microstructural heterogeneities across the disk radius, especially between the disk edge and center. In this study, only the intermediate region (Figures S12 and 7) of the sample is therefore considered.

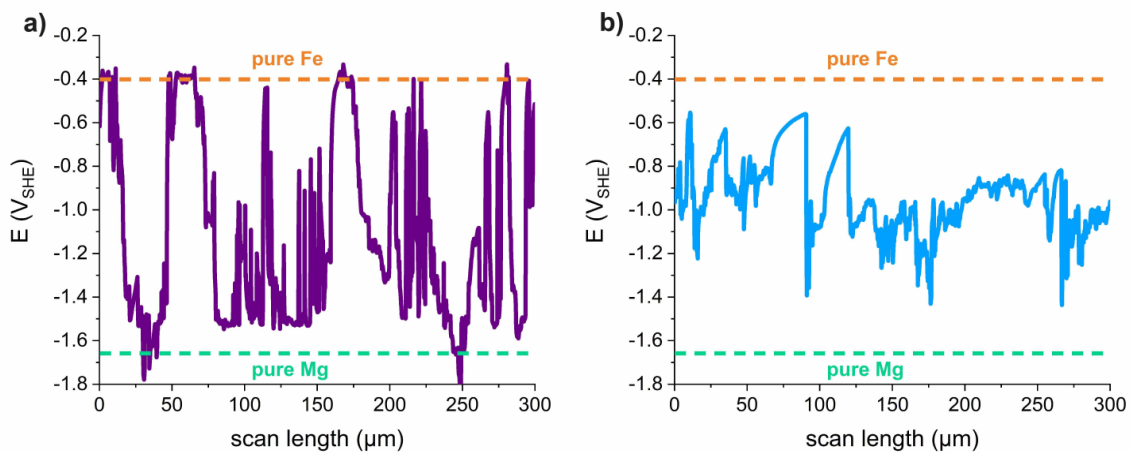

Figure 8: Typical SEN OCP scans for a) coarse and b) nanostructured Mg50-Fe composites, using a 40 nm probe filled with 0.1 M HCl. The SEN line scans were performed with a step size of 200 nm, a scanning rate of 500 nm s$^{-1}$ and a measurement time of 100 ms on each point. The local potential changes displayed are obtained by averaging the potential values over the last 25 ms.

Figure 8 shows that the potential difference between the two phases in the coarse Mg50-Fe composites is higher than that in the nanostructured ones. In particular, for coarse composites, the OCP values vary between ~-1.6 V$_{SHE}$ and -0.4 V$_{SHE}$, depending on whether the nanocapillary is positioned over the Mg phase or the Fe phase respectively. Since these values are close to the OCP values measured on pure Mg and pure Fe, this correlates well with the microstructural analysis of Figure 7 and indicates that the compositions of the two phases and subsequently their reactivity can be considered close to their corresponding pure material. An important galvanic coupling and a large Mg corrosion rate is therefore expected, as confirmed by the macroscopic hydrogen evolution measurements. On the contrary, on nanostructured composites, the Mg phase potential remains above -1.4 $V_{SHE}$, which corresponds to typical OCP values for alloyed Mg. The presence of Fe particles in the Mg phase, too small to be separated from the matrix contribution in these scans, can explain the shift in the electrochemical potential towards more anodic values compared to that of pure Mg. The potential of the Fe phase is also decreased in the nanostructured composite. The galvanic coupling between the two phases is therefore reduced and the overall reactivity is expected to be lower than that of the coarse Mg50-Fe composites. Additional surface modifications related to selective dissolution of Mg and Fe enrichment of the Mg-rich phase or decrease of cathodic activity of the modified Fe phase can add to the decreased electrochemical reactivity observed.

Since the step size of the SEN OCP scans lies roughly in the order of the spacing of the fine lamellae, complementary scanning Kelvin probe force microscopy (SKPFM) was performed. As shown in Figure 9a, on polished surfaces, the Volta potential difference between the two phases is also higher on the coarse composites (around 0.8 V) than on the nanostructured ones (only of 0.5 V). Note that in presence of air humidity, a "Helmholtz part" of electrochemical double layer can already form. The Volta potential values are close to the OCP values measured during SEN scans but obviously are influenced by the presence of this surface air-formed oxide – and possibly smearing. To generate a more defined solid-water interface, short exposure to water after polishing results in an increase in the Volta potential value of the Mg phase in both composites (Figure 9b). The exposure to a thick water layer can induce destabilization/growth of the air-formed surface oxide, in particular on the Mg phase, due to ionic mobility and related galvanic coupling.

Complementary Volta potential measurements were performed on the surfaces after light ion sputtering in high vacuum to remove surface contamination and/or air-formed oxides. This procedure allows to obtain a better characterization of nanostructured composites, but, since after sputtering the surface state is then undefined from a solid-liquid interface perspective, the absolute Volta potential values do not have, in this case, any electrochemical meaning and do not correspond to electrochemical thermodynamic equilibrium values determined in aqueous solutions. As shown in Figure S14, in absence of surface oxide layer formed in contact with water (or at least humidity), the Volta potential difference between the two phases is much more pronounced than on polished surfaces, respectively around 1.3 V on the coarse composites while only of around 0.8 V on the nanostructured ones.

All these findings support the information provided by the SEN OCP scans. Namely, regardless of the surface preparation, i.e., by varying the water contact time and subsequently the oxide formed on the surface, both composites present different surface reactivity, which is intrinsically related to their composite processing conditions and subsequently to their phase spacing and composition. The SEN method has the clear advantage of being able to follow initial stage of oxide chemical stability and corrosion.

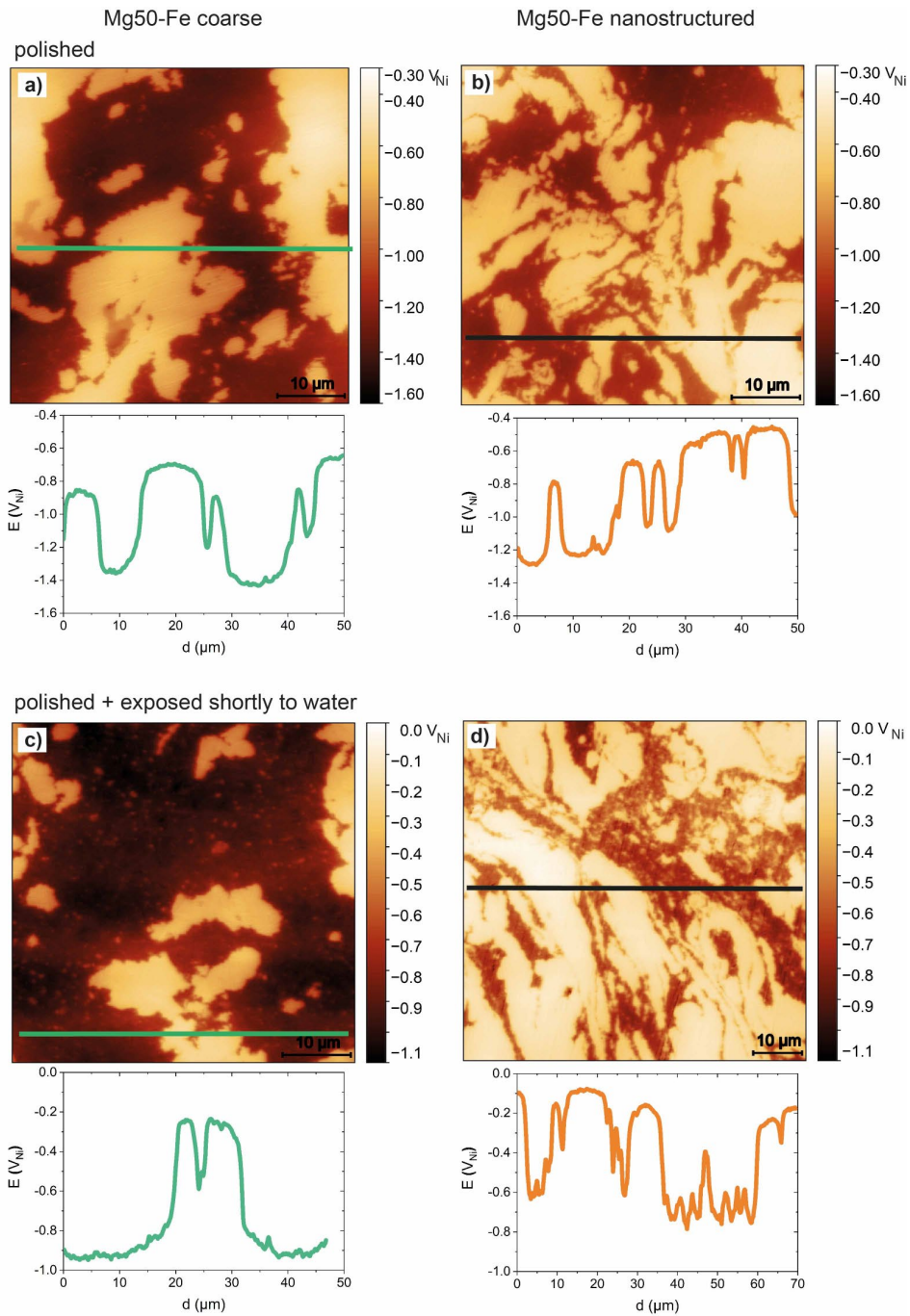

Figure 9: Volta potential measurements of polished a) coarse and b) nanostructured Mg50-Fe composites. c) and d) are the Volta potential measurements on the two composites after short exposure to water. Note that the Volta potentials are expressed versus a Ni reference sample.

SEM observations after SEN OCP scans show that in both composites, the corrosion propagates following the Mg phase. Figure 10 illustrates that corrosion is associated with preferential dissolution of the Mg-rich phase while the Fe-rich phase seems to remain intact even in this aggressive electrolyte. The corrosion attacks moreover seem to propagate further (and laterally) on the nanostructured composite surfaces than on those of the coarse composite, which at first glance contradicts the macroscale corrosion observations, reported in [28]. However, if the presence of Fe particles in the Mg phase of nanostructured composites accelerates the lateral corrosion propagation, it is expected that their refined, nanostructured lamellar microstructure acts as a confined environment. In this constrained space, pH can quickly increase, limiting the in-depth corrosion propagation, similarly to filiform corrosion propagation. The damages are therefore expected to remain superficial in these composites. In comparison, we anticipate that the globular microstructure of the coarse composites, in the means of well-connected micrometer-sized pure Mg phases, favors in-depth corrosion propagation. Complementary post-mortem analysis, such as FIB cross-sectional SEM imaging, not in the focus of this paper on the SEN methods, are however needed to further confirm these suggested propagation mechanisms.

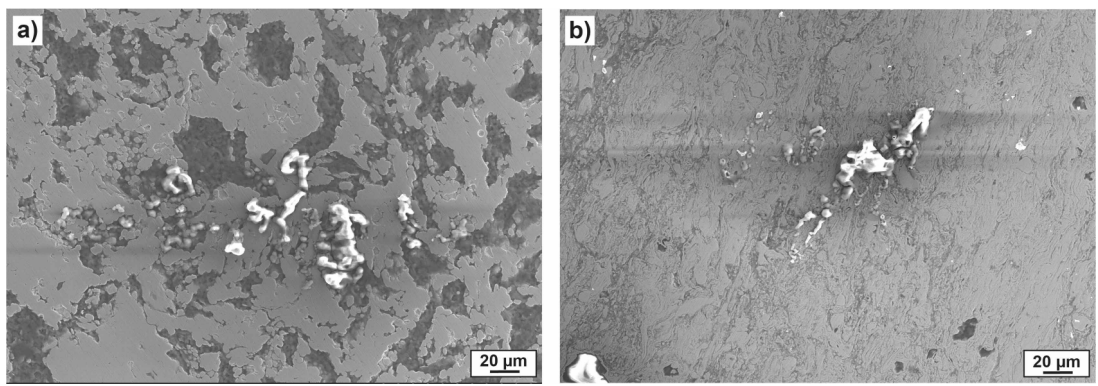

Figure 10: SE SEM image after SEN line scans on a) coarse and b) nanostructured Mg50-Fe composite surface, using a 40 nm probe filled with 0.1 M HCl. The SEN line scans were performed with a step size of 200 nm, a scanning rate of 500 nm s$^{-1}$ and a measurement time of 100 ms on each point.

Local potentiodynamic polarization curves were also performed using the SEN technique, to obtain first kinetic information of the different phases, by individually assessing their behavior. As shown in Figure 11, the potentiodynamic polarization curves obtained on the phases of coarse Mg50-Fe composites exhibit typical active dissolution behavior of either pure Mg (Figure 11, points A-B) or pure Fe (Figure 11, points C-D and comparison with Figure S10). Nevertheless, the kinetic processes are different on both phases. The polarization curves on the Mg-rich phase indicate diffusion-limited dissolution kinetic processes while on the Fe-rich phase, unhindered active dissolution is observed. At the interface between the two phases, an intermediate but still active behavior can be observed (Figure 11, point E). In comparison, on nanostructured Mg50-Fe composites, pure Mg behavior has not been detected, even if sufficiently large area with a single phase could be investigated. Instead, the Mg-rich phase presents an intermediate electrochemical behavior in term of corrosion potential and current, probably induced by the presence of the Fe particles in the Mg matrix. This shows that different dissolution kinetics are evidenced on the Mg-rich phases of coarse and nanostructured composites, which implies different phase stability and thus corrosion processes at the micro- and nanoscale. These findings clearly point to the decisive role of the composite microstructure, i.e. phase spacing and composition, and intrinsically of the processing conditions on the corrosion behavior of Mg-based metallic composites. As discussed before, assessing the effective dissolving area is difficult for this (and any) very localized electrochemical techniques. Starting from a nominal capillary diameter of 40 nm, Figure 6 shows that an exposed droplet size of 200 nm is expected. This would be the situation of the OCP determination mode. For potentiodynamic polarization, it is then expected that this effective area increase even more during active dissolution (Figure S5 and related description). This fact explains the relatively high currents measured and is the reason why we did not corrected the measurement for current densities, because it would give misleading information. The SEN method is able to perform various electrochemical characterizations. Measurement strategy and interpretation should however take into account the specificity of the surface reactivity, for example characterizing differently passive or active surfaces.

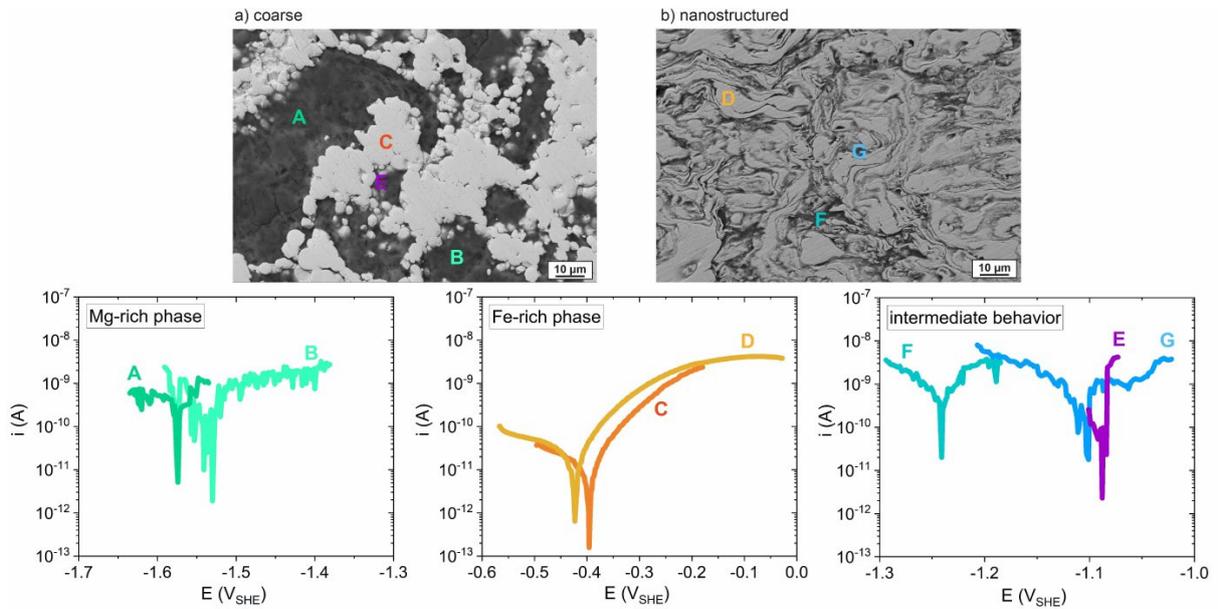

Figure 11: Typical SEN potentiodynamic polarization curves performed on two Mg50-Fe composites with a 40 nm probe filled with 0.1 M HCl at a scan rate of 10 mV s$^{-1}$. Note that the current is not corrected for the area.

## 4. Conclusions

The scanning electrochemical nanocapillary (SEN) technique was successfully applied to characterize the local surface reactivity of active systems, such as Mg-based alloys. Owing to its unprecedented lateral resolution with use of confined corrosion relevant electrolytes, SEN OCP scans offer direct visualization and assessment of the presence, distribution, and nobility of the different intermetallic phases in pure Mg. In particular, high-resolution OCP scans on Mg matrix and single Fe-rich particles confirm that these particles serve as local cathodes while evidencing enhanced surface activation at the interfacial area between the particle and the matrix. But beyond simply identifying microgalvanic coupling between noble particles and Mg matrix, these scans also provide insights about nanoscale galvanic coupling within an intermetallic particle. This information is particularly important in alloys, such as Mg or Al alloys, highly sensitive to localized corrosion for whose initiation, the reactivity of interfacial areas plays a determining role.

Interestingly, although trace amount of Fe drastically affects the corrosion behavior of Mg, a Fe phase fraction of 50 vol.% in Mg-based composites is not necessarily detrimental. Monotonic high pressure torsion (HPT) processing conditions enable the incorporation of Fe particles into the Mg-rich phase, which anodically shifts their electrochemical potential and subsequently reduces the microscale galvanic coupling between the Mg- and Fe-rich phases. Individual assessment of the phase behavior reveals that both, Mg- and Fe-rich phases actively dissolve in chloride-containing aggressive environment. However, depending on the processing conditions, i.e. monotonic or cyclic HPT, very different dissolution kinetics are measured for the Mg-rich phases of the respective composites, which also implies different phase stability and thus corrosion processes at the micro- and nanoscale. These findings clearly point to the determining role of the composite microstructure and intrinsically of the processing conditions on the corrosion behavior of Mg50-Fe composites.

With the example of these composites, we demonstrated the ability of the SEN technique to determine, at the submicrometer scale, surface reactivity and phase stability of complex material systems in reactive electrolytes, information otherwise not easily accessible. By identifying local synergetic effects between dissolution behavior and specific microstructural features, the SEN technique can contribute to refine our mechanistic understanding of the corrosion processes and kinetics of the next generation of light metal alloys and help developing new strategies to improve materials durability and sustainability.

**Acknowledgements**

The authors are grateful to Claudia Cancellieri for performing the XRD measurements and to Dr. Raheleh Partovi-Nia for her contribution in the preliminary study.

This research was financed internally by Empa through instrument development dedicated programs. The transmission electron microscopy facility used in this work received funding from the Austrian Research Promotion Agency (FFG), project known as "3DnanoAnalytics" under contract number FFG-No. 858040."

**Conflict of Interest**

The authors declare no conflict of interest.

**Author contributions**

**Noémie Ott:** Conceptualization, Data Curation, Investigation, Methodology, Validation, Visualization, Writing - Original Draft, Review & Editing. **Aurélien Tournier Fillon:** Investigation, Methodology. **Oliver Renk:** Resources, Investigation, Writing - Review & Editing. **Thomas Suter:** Methodology, Conceptualization. **Thomas Kremmer:** Investigation; Proof-reading. **Stefan Pogatscher:** Resources; Proof-reading. **Patrik Schmutz:** Conceptualization; Project scientific supervision; Resources; Validation; Writing - Review & Editing.

**Data availability**

The data that support the findings of this study are available from the corresponding author upon reasonable request.

**References**

[1] I.J. Polmear, Magnesium alloys and applications, Materials Science and Technology 10 (1994) 1–16. https://doi.org/10.1179/mst.1994.10.1.1.

[2] J.D. Hanawalt, C.E. Nelson, J.A. Peloubet, Corrosion Studies of Magnesium and Its Alloys, Transactions of the AIME 147 (1942) 273–299.

[3] R.E. McNulty, J.D. Hanawalt, Some Corrosion Characteristics of High Purity Magnesium Alloys, Transactions of The Electrochemical Society 81 (1942) 423. https://doi.org/10.1149/1.3071389.

[4] M. Esmaily, J.E. Svensson, S. Fajardo, N. Birbilis, G.S. Frankel, S. Virtanen, R. Arrabal, S. Thomas, L.G. Johansson, Fundamentals and advances in magnesium alloy corrosion, Progress in Materials Science 89 (2017) 92–193. https://doi.org/10.1016/j.pmatsci.2017.04.011.

[5] K. Gusieva, C.H.J. Davies, J.R. Scully, N. Birbilis, Corrosion of magnesium alloys: the role of alloying, International Materials Reviews 60 (2015) 169–194. https://doi.org/10.1179/1743280414Y.0000000046.

[6] A.D. Südholz, N.T. Kirkland, R.G. Buchheit, N. Birbilis, Electrochemical Properties of Intermetallic Phases and Common Impurity Elements in Magnesium Alloys, Electrochemical and Solid-State Letters 14 (2011) C5. https://doi.org/10.1149/1.3523229.

[7] M. Liu, P.J. Uggowitzer, A.V. Nagasekhar, P. Schmutz, M. Easton, G.-L. Song, A. Atrens, Calculated phase diagrams and the corrosion of die-cast Mg–Al alloys, Corrosion Science 51 (2009) 602–619. https://doi.org/10.1016/j.corsci.2008.12.015.

[8] S.K. Woo, B.-C. Suh, H.S. Kim, C.D. Yim, Effect of processing history on corrosion behaviours of high purity Mg, Corrosion Science 184 (2021) 109357. https://doi.org/10.1016/j.corsci.2021.109357.

[9] L. Yang, X. Zhou, S.-M. Liang, R. Schmid-Fetzer, Z. Fan, G. Scamans, J. Robson, G. Thompson, Effect of traces of silicon on the formation of Fe-rich particles in pure magnesium and the corrosion susceptibility of magnesium, Journal of Alloys and Compounds 619 (2015) 396–400. https://doi.org/10.1016/j.jallcom.2014.09.040.

[10] L. Yang, G. Liu, L. Ma, E. Zhang, X. Zhou, G. Thompson, Effect of iron content on the corrosion of pure magnesium: Critical factor for iron tolerance limit, Corrosion Science 139 (2018) 421–429. https://doi.org/10.1016/j.corsci.2018.04.024.

[11] J.F. Löffler, P.J. Uggowitzer, C. Wegmann, M. Becker, H.K. Feichtinger WO2013107644.

[12] J. Hofstetter, E. Martinelli, A.M. Weinberg, M. Becker, B. Mingler, P.J. Uggowitzer, J.F. Löffler, Assessing the degradation performance of ultrahigh-purity magnesium in vitro and in vivo, Corrosion Science 91 (2015) 29–36. https://doi.org/10.1016/j.corsci.2014.09.008.

[13] F. Cao, Z. Shi, J. Hofstetter, P.J. Uggowitzer, G. Song, M. Liu, A. Atrens, Corrosion of ultra-high-purity Mg in 3.5% NaCl solution saturated with Mg(OH)2, Corrosion Science 75 (2013) 78–99. https://doi.org/10.1016/j.corsci.2013.05.018.

[14] C. Wang, Di Mei, G. Wiese, L. Wang, M. Deng, S.V. Lamaka, M.L. Zheludkevich, High rate oxygen reduction reaction during corrosion of ultra-high-purity magnesium, npj Materials Degradation 4 (2020) 42. https://doi.org/10.1038/s41529-020-00146-1.

[15] H. Somekawa, J. Yi, A. Singh, K. Tsuchiya, Microstructural evolution via purity grade of magnesium produced by high pressure torsion, Materials Science and Engineering: A 823 (2021) 141735. https://doi.org/10.1016/j.msea.2021.141735.

[16] D. Ahmadkhaniha, A. Järvenpää, M. Jaskari, M.H. Sohi, A. Zarei-Hanzaki, M. Fedel, F. Deflorian, L.P. Karjalainen, Microstructural modification of pure Mg for improving mechanical and biocorrosion properties, Journal of the Mechanical Behavior of Biomedical Materials 61 (2016) 360–370. https://doi.org/10.1016/j.jmbbm.2016.04.015.

[17] K. Edalati, R. Uehiro, K. Fujiwara, Y. Ikeda, H.-W. Li, X. Sauvage, R.Z. Valiev, E. Akiba, I. Tanaka, Z. Horita, Ultra-severe plastic deformation: Evolution of microstructure, phase transformation and hardness in immiscible magnesium-based systems, Materials Science and Engineering: A 701 (2017) 158–166. https://doi.org/10.1016/j.msea.2017.06.076.

[18] D. Amram, C.A. Schuh, Interplay between thermodynamic and kinetic stabilization mechanisms in nanocrystalline Fe-Mg alloys, Acta Materialia 144 (2018) 447–458. https://doi.org/10.1016/j.actamat.2017.11.014.

[19] K. Edalati, H. Emami, Y. Ikeda, H. Iwaoka, I. Tanaka, E. Akiba, Z. Horita, New nanostructured phases with reversible hydrogen storage capability in immiscible magnesium–zirconium system produced by high-pressure torsion, Acta Materialia 108 (2016) 293–303. https://doi.org/10.1016/j.actamat.2016.02.026.

[20] D.R. Leiva, L.F. Chanchetti, R. Floriano, T.T. Ishikawa, W.J. Botta, Exploring several different routes to produce Mg- based nanomaterials for Hydrogen storage, IOP Conference Series: Materials Science and Engineering 63 (2014) 12115. https://doi.org/10.1088/1757-899x/63/1/012115.

[21] T. Grosdidier, J.J. Fundenberger, J.X. Zou, Y.C. Pan, X.Q. Zeng, Nanostructured Mg based hydrogen storage bulk materials prepared by high pressure torsion consolidation of arc plasma evaporated ultrafine powders, International Journal of Hydrogen Energy 40 (2015) 16985–16991. https://doi.org/10.1016/j.ijhydene.2015.06.159.

[22] G.F. Lima, A.M. Jorge, D.R. Leiva, C.S. Kiminami, C. Bolfarini, W.J. Botta, Severe plastic deformation of Mg-Fe powders to produce bulk hydrides, Journal of Physics: Conference Series 144 (2009) 12015. https://doi.org/10.1088/1742-6596/144/1/012015.

[23] R.B. Figueiredo, T.G. Langdon, Processing Magnesium and Its Alloys by High-Pressure Torsion: An Overview, Advanced Engineering Materials 21 (2019) 1801039. https://doi.org/10.1002/adem.201801039.

[24] Strength, corrosion resistance, and biocompatibility of ultrafine-grained Mg alloys after different modes of severe plastic deformation, 1st ed., 2017.

[25] J. Horky, A. Ghaffar, K. Werbach, B. Mingler, S. Pogatscher, R. Schäublin, D. Setman, P.J. Uggowitzer, J.F. Löffler, M.J. Zehetbauer, Exceptional Strengthening of Biodegradable Mg-Zn-Ca Alloys through High Pressure Torsion and Subsequent Heat Treatment, Materials 12 (2019) 2460.

[26] P. Brunner, F. Brumbauer, E.-M. Steyskal, O. Renk, A.-M. Weinberg, H. Schroettner, R. Würschum, Influence of high-pressure torsion deformation on the corrosion

behaviour of a bioresorbable Mg-based alloy studied by positron annihilation, Biomaterials Science 9 (2021) 4099–4109. https://doi.org/10.1039/D1BM00166C.

[27] W. Li, X. Liu, Y. Zheng, W. Wang, W. Qiao, K.W.K. Yeung, K.M.C. Cheung, S. Guan, O.B. Kulyasova, R.Z. Valiev, In vitro and in vivo studies on ultrafine-grained biodegradable pure Mg, Mg–Ca alloy and Mg–Sr alloy processed by high-pressure torsion, Biomaterials Science 8 (2020) 5071–5087. https://doi.org/10.1039/D0BM00805B.

[28] O. Renk, I. Weißensteiner, M. Cihova, E.-M. Steyskal, N.G. Sommer, M. Tkadletz, S. Pogatscher, P. Schmutz, J. Eckert, P.J. Uggowitzer, R. Pippan, A.M. Weinberg, Mitigating the detrimental effects of galvanic corrosion by nanoscale composite architecture design, npj Materials Degradation 6 (2022) 47. https://doi.org/10.1038/s41529-022-00256-y.

[29] V. Shkirskiy, P. Volovitch, V. Vivier, Development of quantitative Local Electrochemical Impedance Mapping: an efficient tool for the evaluation of delamination kinetics, Electrochimica Acta 235 (2017) 442–452. https://doi.org/10.1016/j.electacta.2017.03.076.

[30] G.L. Song, A. Atrens, Corrosion Mechanisms of Magnesium Alloys, Advanced Engineering Materials 1 (1999) 11–33. https://doi.org/10.1002/(SICI)1527-2648(199909)1:1<11:AID-ADEM11>3.0.CO;2-N.

[31] S.V. Lamaka, O.V. Karavai, A.C. Bastos, M.L. Zheludkevich, M.G.S. Ferreira, Monitoring local spatial distribution of Mg2+, pH and ionic currents, Electrochemistry Communications 10 (2008) 259–262. https://doi.org/10.1016/j.elecom.2007.12.003.

[32] G. Williams, H. Neil McMurray, Localized Corrosion of Magnesium in Chloride-Containing Electrolyte Studied by a Scanning Vibrating Electrode Technique, Journal of The Electrochemical Society 155 (2008) C340. https://doi.org/10.1149/1.2918900.

[33] G. Williams, R. Grace, Chloride-induced filiform corrosion of organic-coated magnesium, Electrochimica Acta 56 (2011) 1894–1903. https://doi.org/10.1016/j.electacta.2010.09.005.

[34] K. Kondoh, N. Nakanishi, R. Takei, H. Fukuda, J. Umeda, Evaluation of Initial Corrosion Phenomenon of Magnesium Alloys by SKPFM, Materials Science Forum 690 (2011) 397–400. https://doi.org/10.4028/www.scientific.net/MSF.690.397.

[35] G.-L. Song, Z. Xu, Crystal orientation and electrochemical corrosion of polycrystalline Mg, Corrosion Science 63 (2012) 100–112. https://doi.org/10.1016/j.corsci.2012.05.019.

[36] J. Izquierdo, L. Nagy, I. Bitter, R.M. Souto, G. Nagy, Potentiometric scanning electrochemical microscopy for the local characterization of the electrochemical behaviour of magnesium-based materials, Electrochimica Acta 87 (2013) 283–293. https://doi.org/10.1016/j.electacta.2012.09.029.

[37] G. Williams, N. Birbilis, H.N. McMurray, The source of hydrogen evolved from a magnesium anode, Electrochemistry Communications 36 (2013) 1–5. https://doi.org/10.1016/j.elecom.2013.08.023.

[38] P. Dauphin-Ducharme, J. Mauzeroll, Surface Analytical Methods Applied to Magnesium Corrosion, Analytical Chemistry 87 (2015) 7499–7509. https://doi.org/10.1021/ac504576g.

[39] S.V. Lamaka, J. Gonzalez, Di Mei, F. Feyerabend, R. Willumeit-Römer, M.L. Zheludkevich, Local pH and Its Evolution Near Mg Alloy Surfaces Exposed to Simulated Body Fluids, Advanced Materials Interfaces 5 (2018) 1800169. https://doi.org/10.1002/admi.201800169.

[40] G. Baril, C. Blanc, M. Keddam, N. Pébère, Local Electrochemical Impedance Spectroscopy Applied to the Corrosion Behavior of an AZ91 Magnesium Alloy, Journal of The Electrochemical Society 150 (2003) B488. https://doi.org/10.1149/1.1602080.

[41] G. Galicia, N. Pébère, B. Tribollet, V. Vivier, Local and global electrochemical impedances applied to the corrosion behaviour of an AZ91 magnesium alloy, Corrosion Science 51 (2009) 1789–1794. https://doi.org/10.1016/j.corsci.2009.05.005.

[42] S. Fajardo, C.F. Glover, G. Williams, G.S. Frankel, The Source of Anodic Hydrogen Evolution on Ultra High Purity Magnesium, Electrochimica Acta 212 (2016) 510–521. https://doi.org/10.1016/j.electacta.2016.07.018.

[43] S.H. Salleh, S. Thomas, J.A. Yuwono, K. Venkatesan, N. Birbilis, Enhanced hydrogen evolution on Mg (OH)2 covered Mg surfaces, Electrochimica Acta 161 (2015) 144–152. https://doi.org/10.1016/j.electacta.2015.02.079.

[44] P. Dauphin-Ducharme, R. Matthew Asmussen, U.M. Tefashe, M. Danaie, W. Jeffrey Binns, P. Jakupi, G.A. Botton, D.W. Shoesmith, J. Mauzeroll, Local Hydrogen Fluxes Correlated to Microstructural Features of a Corroding Sand Cast AM50 Magnesium Alloy, Journal of The Electrochemical Society 161 (2014) C557-C564. https://doi.org/10.1149/2.0571412jes.

[45] N.A. Payne, L.I. Stephens, J. Mauzeroll, The Application of Scanning Electrochemical Microscopy to Corrosion Research, Corrosion 73 (2017) 759–780. https://doi.org/10.5006/2354.

[46] G. Wittstock, M. Burchardt, S.E. Pust, Y. Shen, C. Zhao, Scanning Electrochemical Microscopy for Direct Imaging of Reaction Rates, Angewandte Chemie International Edition 46 (2007) 1584–1617. https://doi.org/10.1002/anie.200602750.

[47] M. Jönsson, D. Thierry, N. LeBozec, The influence of microstructure on the corrosion behaviour of AZ91D studied by scanning Kelvin probe force microscopy and scanning Kelvin probe, Corrosion Science 48 (2006) 1193–1208. https://doi.org/10.1016/j.corsci.2005.05.008.

[48] M.F. Hurley, C.M. Efaw, P.H. Davis, J.R. Croteau, E. Graugnard, N. Birbilis, Volta Potentials Measured by Scanning Kelvin Probe Force Microscopy as Relevant to

Corrosion of Magnesium Alloys, Corrosion 71 (2015) 160–170. https://doi.org/10.5006/1432.

[49] L. Eng, E. Wirth, T. Suter, H. Böhni, Non-contact feedback for scanning capillary microscopy, Electrochimica Acta 43 (1998) 3029–3033. https://doi.org/10.1016/S0013-4686(98)00043-7.

[50] L. Staemmler, T. Suter, H. Böhni, Glass capillaries as a tool in nanoelectrochemical deposition, Electrochemical and Solid-State Letters 5 (2002) C61-C63. https://doi.org/10.1149/1.1473257.

[51] L. Staemmler, T. Suter, H. Böhni, Nanolithography by means of an electrochemical scanning capillary microscope, Journal of The Electrochemical Society 151 (2004) G734-G739. https://doi.org/10.1149/1.1803834.

[52] M.W. Kapp, O. Renk, T. Leitner, P. Ghosh, B. Yang, R. Pippan, Cyclically induced grain growth within shear bands investigated in UFG Ni by cyclic high pressure torsion, Journal of Materials Research 32 (2017) 4317–4326. https://doi.org/10.1557/jmr.2017.273.

[53] P. Schmutz, G.S. Frankel, Characterization of AA2024-T3 by Scanning Kelvin Probe Force Microscopy, Journal of The Electrochemical Society 145 (1998) 2285–2295. https://doi.org/10.1149/1.1838633.

[54] T. Suter, H. Böhni, Local electrochemical methods to study corrosion processes on a molecular level, Zairyo to Kankyo/ Corrosion Engineering 50 (2001) 349–363. https://doi.org/10.3323/jcorr1991.50.349.

[55] T. Suter, H. Böhni, A new microelectrochemical method to study pit initiation on stainless steels, Electrochimica Acta 42 (1997) 3275–3280. https://doi.org/10.1016/S0013-4686(70)01783-8.

[56] H. Böhni, T. Suter, A. Schreyer, Micro- and nanotechniques to study localized corrosion, Electrochimica Acta 40 (1995) 1361–1368. https://doi.org/10.1016/0013-4686(95)00072-M.

[57] T. Suter, Mikroelektronische Untersuchungen bei austenitischen "rostfreien" Stählen, 1997.

[58] M. Taheri, J.R. Kish, N. Birbilis, M. Danaie, E.A. McNally, J.R. McDermid, Towards a Physical Description for the Origin of Enhanced Catalytic Activity of Corroding Magnesium Surfaces, Electrochimica Acta 116 (2014) 396–403. https://doi.org/10.1016/j.electacta.2013.11.086.

[59] T. Cain, S.B. Madden, N. Birbilis, J.R. Scully, Evidence of the Enrichment of Transition Metal Elements on Corroding Magnesium Surfaces Using Rutherford Backscattering Spectrometry, Journal of The Electrochemical Society 162 (2015) C228-C237. https://doi.org/10.1149/2.0541506jes.

[60] L. Yang, X. Zhou, M. Curioni, S. Pawar, H. Liu, Z. Fan, G. Scamans, G. Thompson, Corrosion Behavior of Pure Magnesium with Low Iron Content in 3.5 wt% NaCl Solution, Journal of The Electrochemical Society 162 (2015) C362-C368. https://doi.org/10.1149/2.1041507jes.

[61] D. Höche, C. Blawert, S.V. Lamaka, N. Scharnagl, C. Mendis, M.L. Zheludkevich, The effect of iron re-deposition on the corrosion of impurity-containing magnesium, Physical Chemistry Chemical Physics 18 (2016) 1279–1291. https://doi.org/10.1039/C5CP05577F.

*Supplementary Information*

# Local electrochemical characterization of active Mg-Fe materials – from pure Mg to Mg50-Fe composites

Noémie Ott[a,ǂ*], Aurélien Tournier Fillon[a], Oliver Renk[b], Thomas Kremmer[c], Stefan Pogatscher[c], Thomas Suter[a,†], Patrik Schmutz[a,*]

[a] Laboratory for Joining Technologies and Corrosion, Empa – Swiss Federal Laboratories for Materials Science and Technology, Überlandstrasse 129, 8600 Dübendorf, Switzerland

[b] Department Materials Science, Montanuniversität Leoben, Roseggerstraße 12, 8700 Leoben, Austria

[c] Chair of Nonferrous Metallurgy, Montanuniversität Leoben, Franz-Josef-Strasse 18, 8700 Leoben, Austria

* Corresponding authors: noemie.ott@ost.ch (N. Ott), patrik.schmutz@empa.ch (P. Schmutz)

† Deceased on April 25th, 2020 (Dr. T. Suter)

ǂ Current address (N. Ott): Institute for Microtechnology and Photonics, OST, Werdenbergstrasse 4, 9471 Buchs, Switzerland

1. **Scanning electrochemical nanocapillary (SEN) method**
   a. **Probes**

The probes with a nominal diameter of 40 nm used in this study are obtained by pulling borosilicate glass capillaries (OD: 1.0 mm, ID: 0.58 mm, with filament, Science Products) in a one-stage process with a laser-based micropipette puller (P-2000, Sutter Instruments). Figure S1 shows the STEM-in-SEM imaging of such a tip.

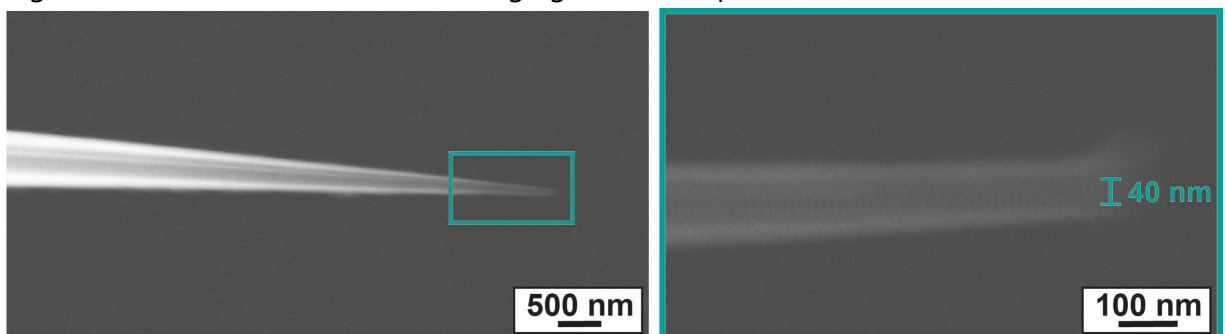

**Figure S1:** STEM-in-SEM image of the capillary probe tip.

Before each scan, the resonance frequency of the oscillating probe is determined but typically lies in the 40 – 80 kHz range (Figure S2). The Q-factor of the probe is determined as the ratio of the resonance frequency to the peak full width at half maximum (FWHM). Only probes with a Q-factor above 100 are considered.

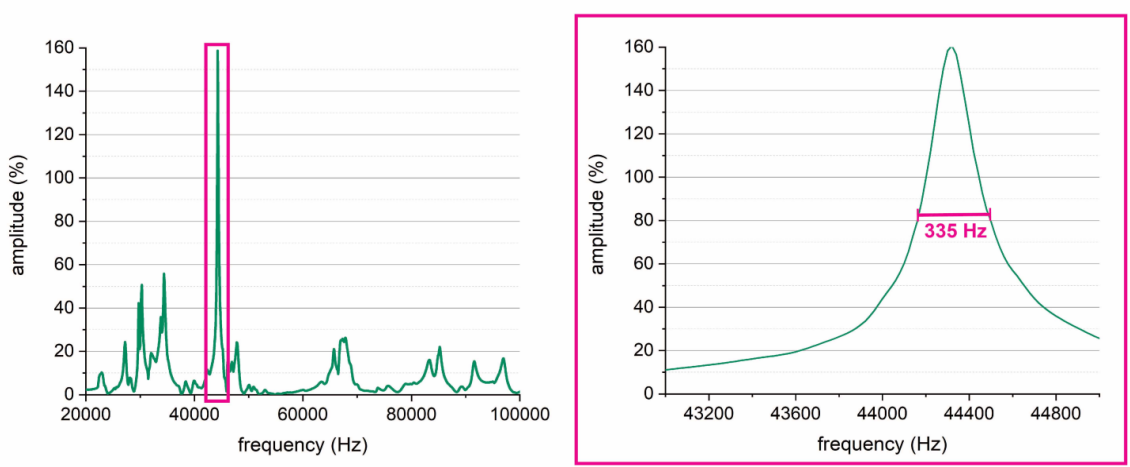

**Figure S2:** Typical example of the frequency-dependent response of the oscillating capillary probe. In this case, the probe resonance frequency is 44.320 kHz and its Q-factor is 130 %.

   b. **Topographic information – calibration**

A TGZ03 calibration grid (3 µm pitch, height: 496 nm, MikroMasch) was used to prove the SEN ability to track small and/or abrupt topographic changes. Mapping (100 µm x 50 µm) was performed with a step size of 200 nm and a scanning rate of 500 nm s$^{-1}$, using a 40 nm capillary probe. The probe stayed for 200 ms on a point. The height value displayed in Figure S3 is obtained by averaging the piezo height displacement over the last 50 ms of the contact time.

For these scans, the nanocapillary was filled with MilliQ water (Merck Millipore, 18.2 MΩ cm). Figure S3 clearly demonstrates that the expected topography is recovered. Note that the acquisition of such a map took on purpose more than 24 h without any loss in tracking capability. This proves the stability of the feedback process.

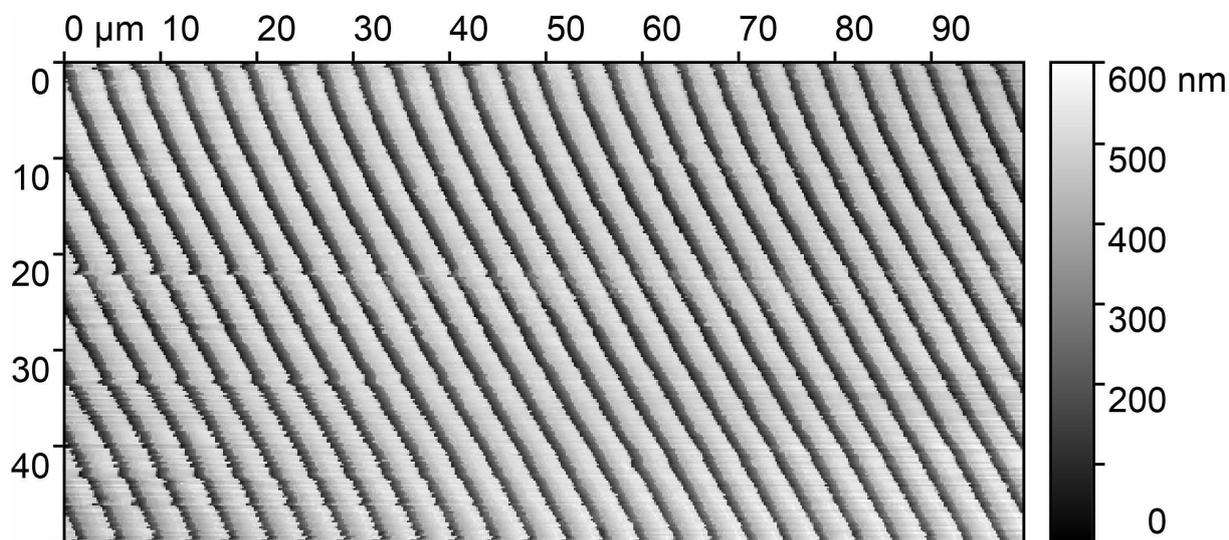

**Figure S3:** Mapping of TGZ03 calibration grid (3 µm pitch, height: 496 nm, MikroMasch) with a 40 nm probe filled with MilliQ water. This shows the ability of the SEN setup to track even small topographic changes over long time (> 24 h acquisition time).

### c. Tracking potential changes – calibration

The PFKPFM-SMPL reference sample (Bruker) was used as reference to demonstrate the SEN ability to measure microscale potential differences. Line scans were performed at open-circuit potential (OCP) with a step size of 300 nm, a scanning rate of 500 nm s$^{-1}$ and a measurement duration of 200 ms on a point, using a 40 nm probe filled with 0.001 M HCl at pH 3.0. Complementary scanning Kelvin probe force microscopy (AFM/SKPFM) was performed in air.

Figure S4 confirms the good topographic and potential tracking ability of the SEN instrument. Note that in this figure, the SEN topographic data were not leveled. A tilt of 400 nm over a scan length of 100 µm can therefore be observed in Figure S4 c). Nevertheless, the SEN good height control allows retrieving comparable topographic information to AFM imaging, despite slightly lower resolution. The potential values measured on both metals are consistent with the values expected at such pH for thin films of Au and Al. In HCl at pH 3.0, Al is near the active-to-passive transition but the 200 ms measurement duration is not sufficient to fully activate the surface. The open-circuit potential (OCP) measured indicates the formation of a defective but yet protective oxide layer on Al while the SKPFM potential measured in air corresponds to an oxidized, carbon contaminated surface. Surface cleaning (typically by a sputtering process to avoid microscale interactions between materials) would be required for a better assessment of the surface potential.

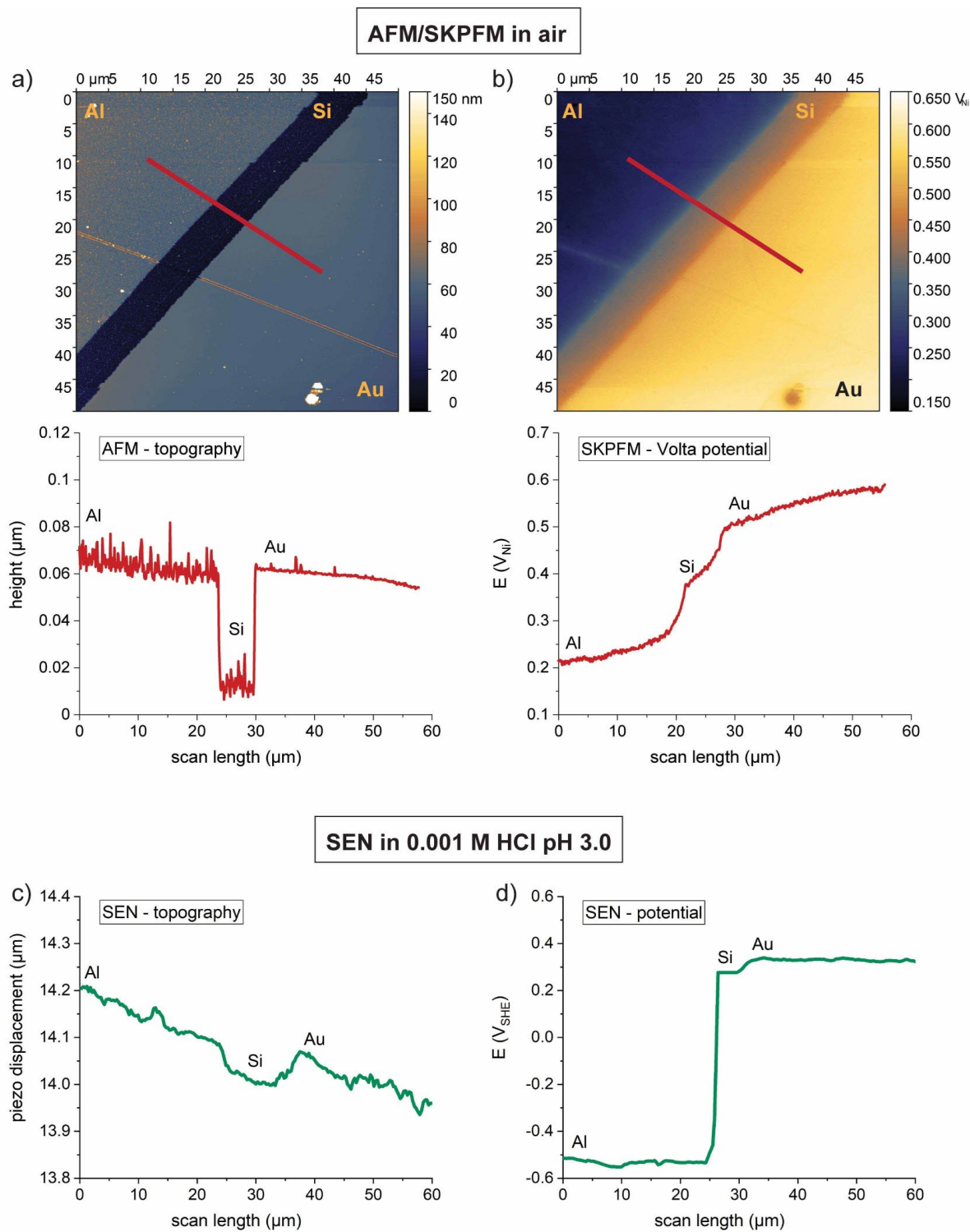

**Figure S4:** a) AFM and b) SKPMF imaging of the PFKPFM-SMPL reference sample in air. Note that the Volta potentials are expressed versus a Ni reference. Corresponding c) topographic and d) potential scans obtained by SEN measurements with a 40 nm probe filled with 0.001 M HCl at pH 3.0.

### d. Effective exposed area

An important aspect of the SEN measurement is the increasing exposed area upon longer electrolyte contact. When measuring large areas in macroscale electrochemistry, the influence on measured currents of geometrical changes related to corrosion is negligible. Using capillaries with ultra-small nominal diameter of 40 nm results in a totally different situation that needs to be addressed (Figure S5). The effective exposed surface can obviously increase drastically as function of exposure time for active dissolution processes and can even be order of magnitude larger when potentiodynamic polarization are performed. For this reason, some compromise needs to be taken depending on the information aimed at, but longer exposition times can also benefit in allowing to investigate corrosion propagation mechanism and bringing the current range in an "easily" measurable range.

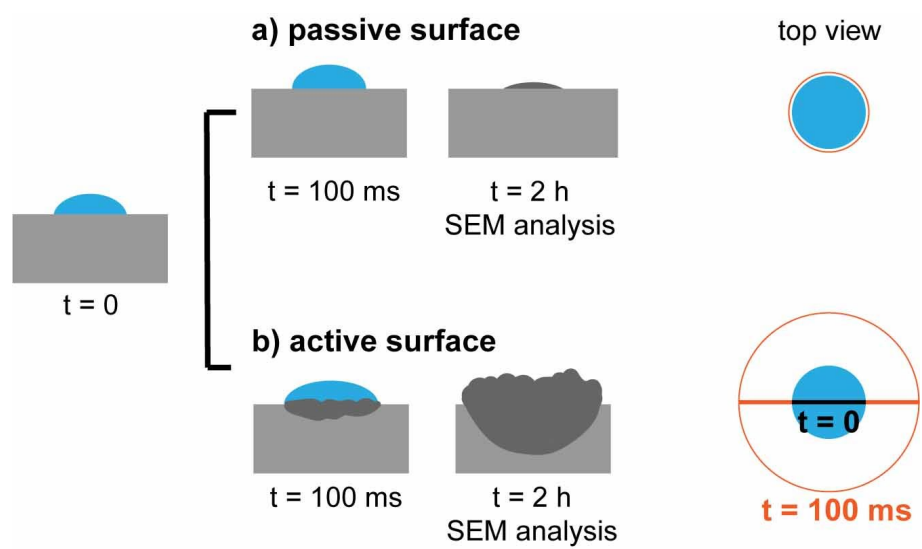

**Figure S5:** Schematics showing the evolution of the exposed area on a) passive and b) active surfaces. The effective surface is not expected to change on passive surfaces while on active surfaces, it is expected to drastically increases as soon as active dissolution starts, which can lead to an estimation mismatch. Therefore, the current was not corrected for the area in SEN-based measurements.

## 2. Microstructural characterization of pure Mg materials

**a) Mg CP**

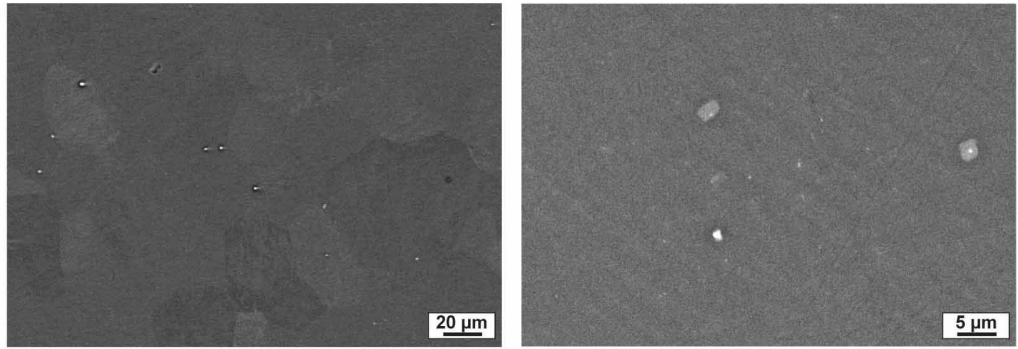

**b) Mg HP**

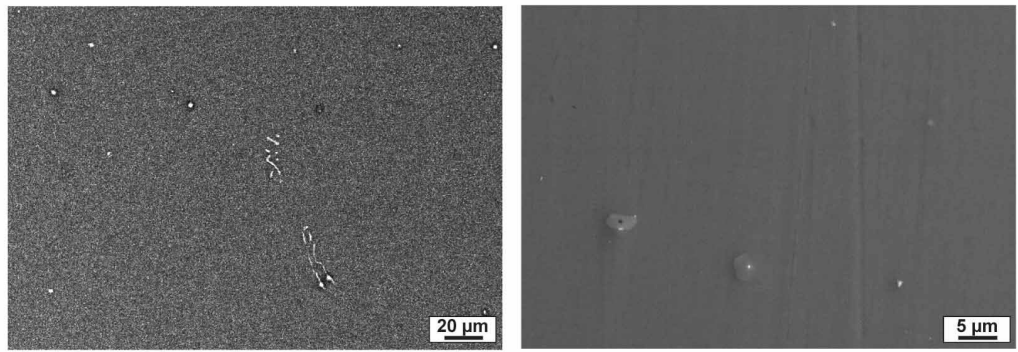

**c) Mg HP HT**

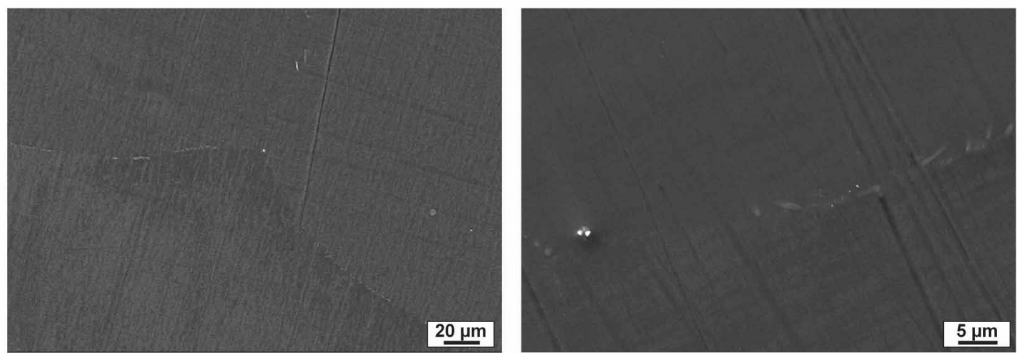

**Figure S6:** Microstructure of the pure Mg grades studied. a) commercially pure Mg (Mg CP); high purity Mg b) as-produced (Mg HP) and c) after heat-treatment for 24 h at 525 °C under Ar atmosphere (Mg HP HT).

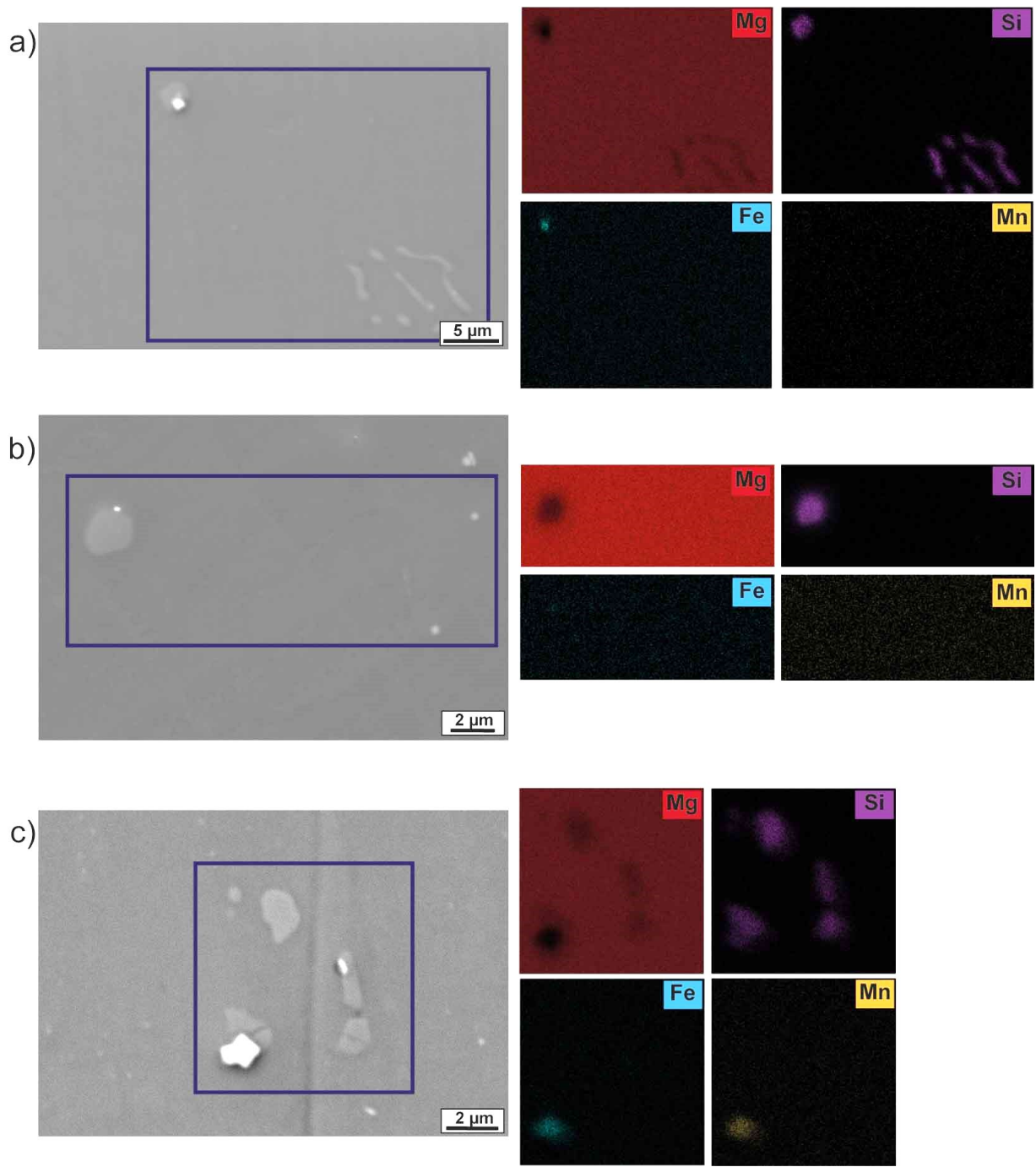

**Figure S7:** SEM images and corresponding energy-dispersive X-ray spectroscopy (EDX) maps of particles found in high purity Mg a) as-produced (HP) and b) after heat-treatment for 24 h at 525 °C under Ar atmosphere (HP HT) and c) in commercially pure Mg materials (CP)

## 3. Dissolution behavior of pure Mg grades

The macroscale corrosion behavior of pure Mg was estimated by volumetric hydrogen collection measurements. Pure Mg samples, i.e. high purity Mg as-produced (HP), Mg HP after heat-treatment (HP-HT) for 24 h at 525 °C and commercially pure Mg (CP), were each placed in a beaker containing 1.0 M NaCl at an initial pH value of 6.0. A measuring cylinder also filled with this unbuffered NaCl solution was placed directly on top of the sample in order to collect all the hydrogen produced by the dissolving samples. The volume of $H_2$ produced is proportional to Mg dissolution and thus, can be directly used to derive the Mg corrosion rate. As would be expected, the hydrogen evolution rate on CP Mg is about 70 times higher than on HP Mg while heat-treatment of HP Mg leads to a slightly more active surface than in the as-produced condition (Figure S8a). This confirms that not only the impurity content but also the alloy processing parameters play a significant role on the corrosion behavior of pure Mg.

Complementary immersion tests followed by element-specific analysis were performed. HP and CP Mg samples were immersed in unbuffered 0.15 M NaCl at an initial pH value of 5.2 (Suprapur grade). Aliquots were sampled every 10 min during 180 min and subsequently diluted five or ten times with a 1 % nitric acid electrolyte (Suprapur grade). The analysis of the solutions was conducted using an Agilent 7500ce inductively coupled plasma mass spectrometer (ICPMS). The calibration was performed using matrix-matched blank solution and multi-element standard solutions in a concentration range from 0.05 µg L$^{-1}$ to 5000 µg L$^{-1}$. An internal standard (IS) solution was added online to correct for non-spectral interferences. Only concentrations above the background equivalent concentration (BEC) were taken into account. Typical BEC values for the elements of interest are: $^{25}$Mg=0.46 µg L$^{-1}$, $^{55}$Mn=0.011 µg L$^{-1}$, $^{56}$Fe=0.16 µg L$^{-1}$ and $^{68}$Zn=0.16 µg L$^{-1}$. The performance of the ICPMS (accuracy and signal drift) was periodically checked through calibration validation using NIST SRM 1643 certified reference samples and quality control samples. The recovery was within 5% each time, which validates the method used.

After normalization by the IS intensity and subtraction of the corresponding blank value, the measured concentrations were corrected for the dilution, which occurs each time an aliquot is sampled out and replaced by fresh solution. The "real" amount of a dissolved element at each step is therefore the sum of the actual measured element concentration and the amount, which was previously sampled. The dissolved amount of each element was corrected for the initial weight of the sample and is reported as µg g$^{-1}$. The immersion tests were repeated three times for both purity grades.

Note that the experiments were conducted in unbuffered NaCl solutions. As shown in Figure S9, the pH quickly increases towards alkaline pH values, allowing the formation of corrosion products at the surface. Macroscopic observations of the samples after immersion (Figure S8) show the presence of dark filiform-like tracks of corrosion products on the surface of CP Mg compared to a thinner oxyhydroxide layer formed on HP Mg surface.

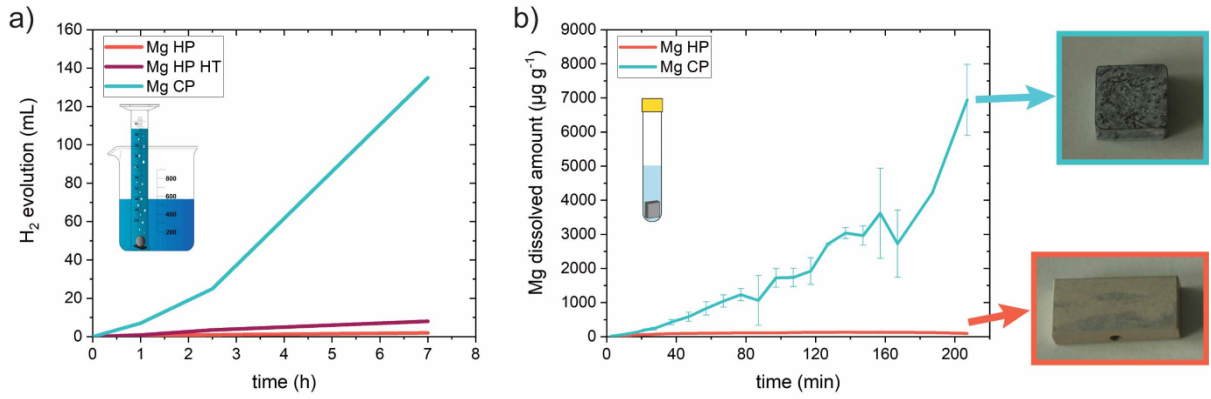

**Figure S8:** Effect of impurities on the corrosion rates of pure Mg. a) Hydrogen evolution and b) amount of dissolved Mg during immersion of high purity (HP, HP-HT) and commercially pure (CP) Mg alloys in unbuffered NaCl solutions.

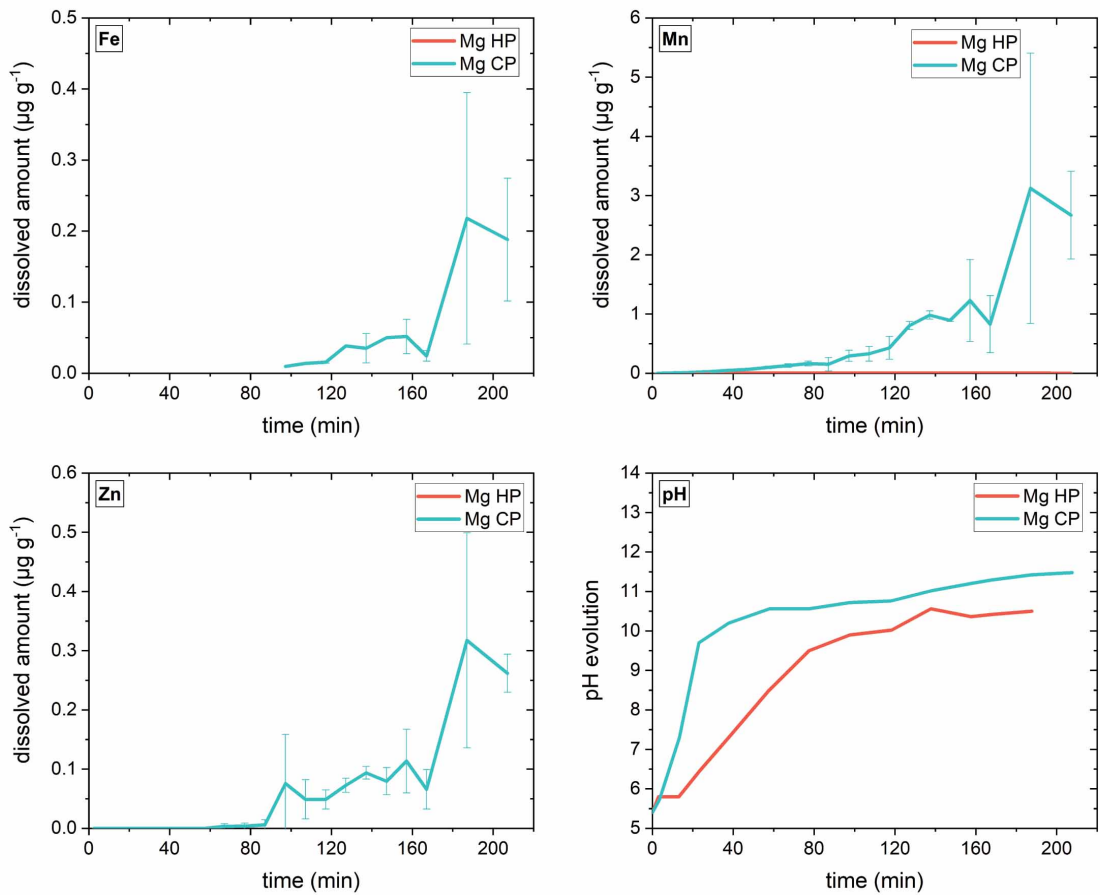

**Figure S9:** Dissolved amount of trace-elements during immersion of high purity (HP) and commercially pure (CP) Mg in unbuffered 0.1 M NaCl and the corresponding pH evolution.

## 4. SEN measurements on pure Fe

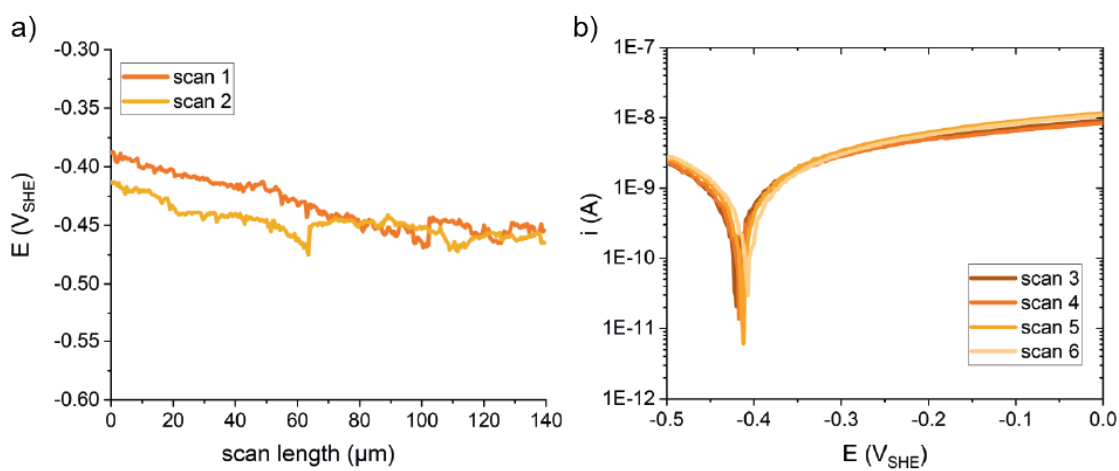

**Figure S10:** SEN measurements on Fe 99.5% with a 40 nm probe filled with 0.1 M HCl. a) potential scans with a measurement duration of 100 ms per point and b) potentiodynamic polarization curves at a scan rate of 10 mV s$^{-1}$. Note that the current is not corrected for the area.

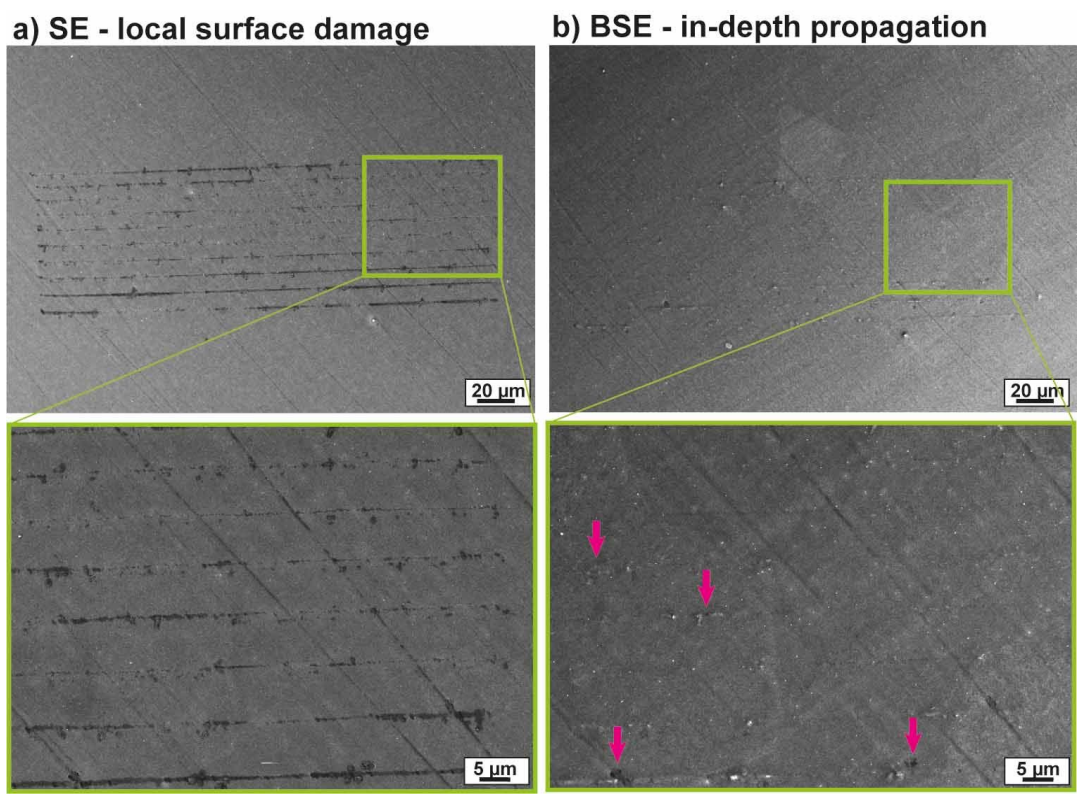

**Figure S11:** a) Secondary electron (SE) and b) Backscattered electron (BSE) images after SEN line scans on Mg CP surfaces, using a 40 nm probe filled with 0.1 M HCl. The SEN line scans were performed with a step size of 400 nm, a scanning rate of 500 nm s$^{-1}$ and a measurement time of 100 ms on each point. The distance between each line scan is 5 µm.

## 5. Mg50-Fe composites

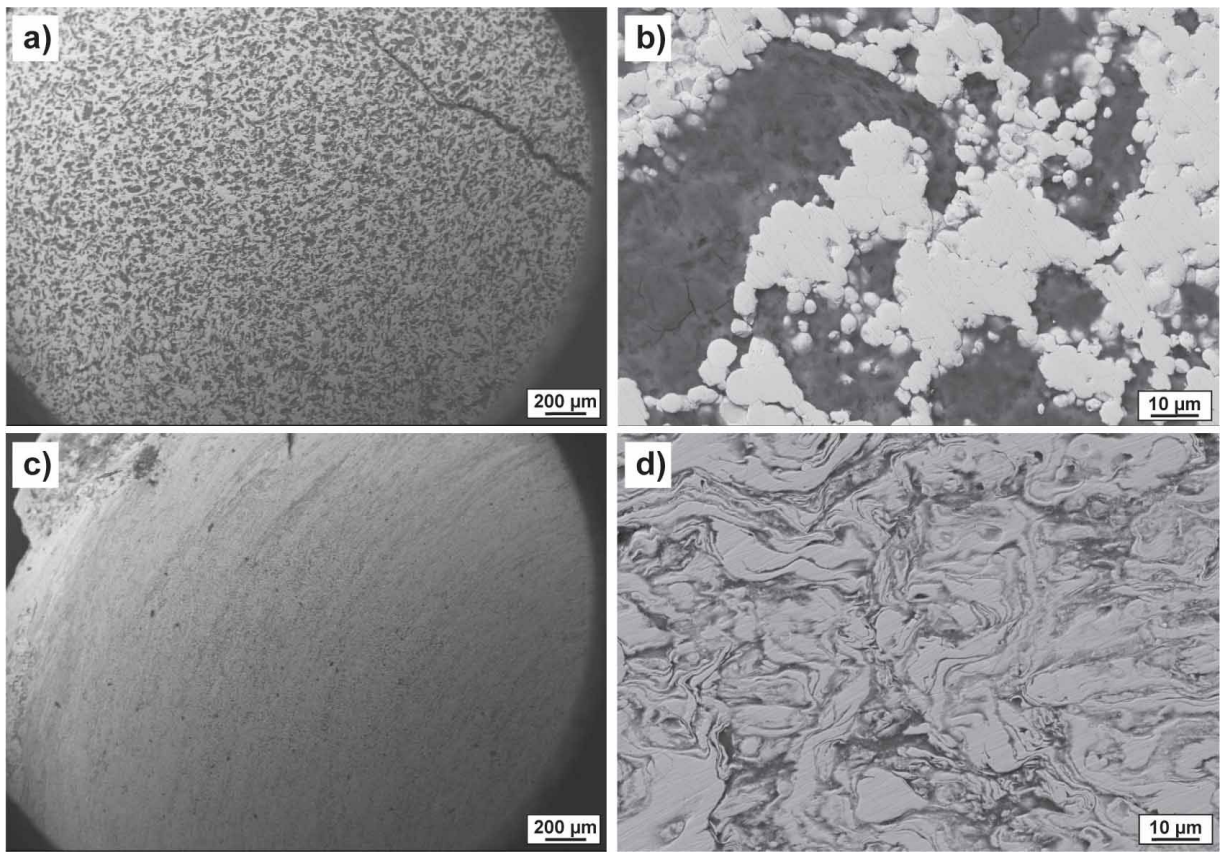

**Figure S12:** SEM images of Mg50-Fe composites produced by a) and b) cyclic HPT at room temperature (coarse) and c) and d) monotonic HPT at 300 °C (nanostructured).

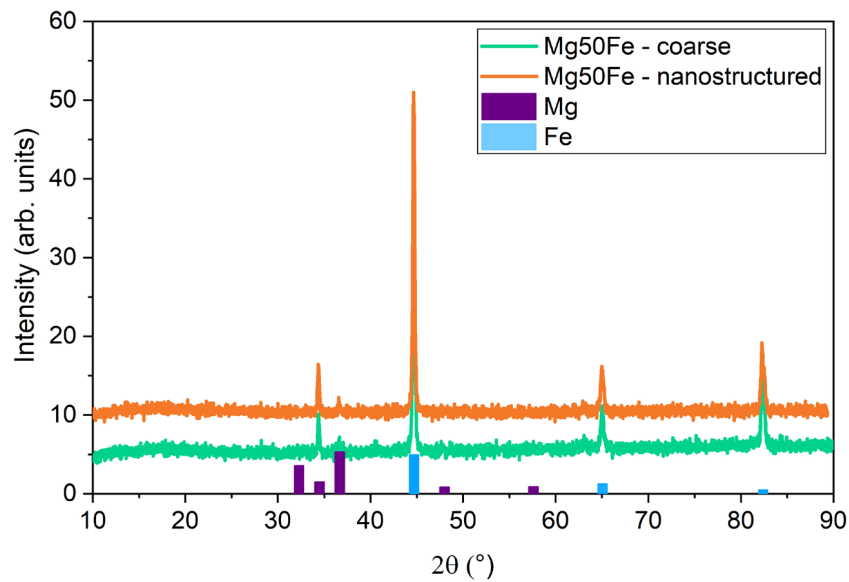

**Figure S13:** XRD scans for the two Mg50-Fe composites and corresponding peak positions for Mg (hcp, collection code: 29734) and Fe (bcc, collection code: 52258). These reference powder diffraction patterns were taken from the online ICSD database.

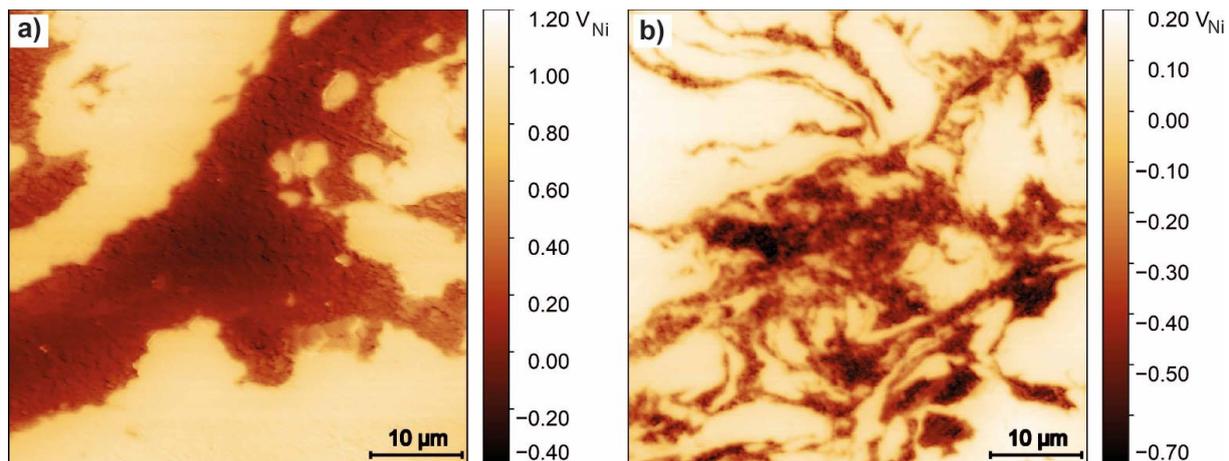

**Figure S14:** SKPFM based Volta potential measurements on sputtered a) coarse and b) nanostructured Mg50-Fe composite surfaces. Note that the Volta potentials are expressed versus a Ni reference sample.